\documentclass[12pt]{article}
\usepackage{amssymb}
\usepackage{amsmath}
\usepackage{graphicx}
\usepackage{nicefrac}
\title{Primordial black hole formation in $k$-inflation models}
\author{
	Neven Bili\'c\thanks{bilic@irb.hr, corresponding author}
	\\
	{\it Division of Theoretical Physics, Rudjer Bo\v{s}kovi\'{c} Institute,}
	\\ {\it Zagreb, Croatia}
	\vspace{0.1in}
	\and
	Dragoljub D.\ Dimitrijevi\'{c}\thanks{ddrag@pmf.ni.ac.rs}, 
	Goran S.\ Djordjevi\'{c}\thanks{gorandj@junis.ni.ac.rs}, Milan Milo\v{s}evi\'{c}\thanks{milan.milosevic@pmf.edu.rs}
	\\
	{\it Department of Physics,
		University of Ni\v{s},  Serbia}
	\vspace{0.1in}
	\and
	Marko Stojanovi\'{c}\thanks{marko.stojanovic@pmf.edu.rs} 
	\\
	{\it Faculty of Medicine, University of Ni\v{s}, Serbia}
}

\begin{document}
\maketitle
\flushbottom

	\thispagestyle{empty}
	
	\vspace{0.3cm}

\abstract{
The local primordial density fluctuations caused by quantum vacuum fluctuations during inflation grow into stars and galaxies in the late universe and, if they are large enough, also produce primordial black holes.	We study the formation of the primordial black holes in $k$-essence inflation models with a potential characterized by an inflection point. The background and perturbation equations are integrated numerically for two specific models. Using the critical collapse and peaks formalism, we calculate the abundance of primordial black holes today.
}

\section{Introduction}

According to the original idea of Zel'dovich and Novikov \cite{zeldovich_novikov},
highly overdense regions of inhomogeneities in the early universe could 
gravitationally collapse to form black holes (BHs) 
\cite{ hawking, carr_hawking, carr, chapline},
the so-called primordial black holes
(PBHs) \cite{chapline,carr_kuhnel}.
It has been argued that PBHs may represent a non-negligible component of cold dark matter that provides
the seed for the supermassive BHs that populate the universe.

To generate sufficiently large primordial fluctuations during inflation and form superdense regions that can collapse and become PBHs during the radiation-dominated era, the inflationary scalar perturbations need to be amplified by several orders of magnitude on small scales, 
i.e., for wavenumbers $q$ that are large compared with the pivot scale $q_{\rm CMB}$ of the cosmic microwave background (CMB). 
These scales correspond to the late stages of inflation and are well above the values required to be consistent with CMB observations.

An overdense region collapses and forms a PBH when the local value of the density contrast $\delta=\delta\rho/\rho$ is above the so-called collapse threshold density contrast $\delta_c$.
With the help of Newton's gravity and Jeans' radius $R_{\rm J}\sim c_{\rm s}/H$, it was established that PBH forms when the density contrast at the entry of a certain comoving scale $q$ into the acoustic horizon
(i.e., when $c_{\rm s}q=aH$) exceeds a threshold value
$\delta_{\rm c}\sim c_{\rm s}^2$ \cite{Green:2020jor}. Here, $c_{\rm s}$ is the speed of sound equal to $1/\sqrt3$ in the case of radiation dominance.
The mass of the resulting PBH is approximately equal to the mass within the horizon at the time of formation.
Numerical checks for different primordial perturbation profiles show that during radiation dominance, the value for the threshold density contrast $\delta_c$ is in the range from  $0.4$ to $2/3$ \cite{musco,germani,escriva}.
 
Depending on the mass, PBHs trigger different observational
signals. PBHs lighter than $\sim 10^{15}$ g are already evaporated today by Hawking radiation and do not exist in today's
Universe. Nevertheless, they leave some traces that allow us to investigate how many PBHs might have existed in the early Universe. The abundance of PBHs lighter than $10^{15}$ g, for example, is constrained by the Big Bang nucleosynthesis since the evaporation of high-energy particles changes the abundance of light elements \cite{carr2}.

PBHs heavier than
$10^{15}$ g have not yet lost their mass
significantly by evaporation and remain in the present
Universe. They produce various distinct signals at present
time, such as
\begin{itemize}
	\item 
	gravitational lensing,
	\item
	dynamical effects on baryonic matter,
	\item
	radiation emanating from the accreting matter into PBHs.
\end{itemize}
The choice of the physical process needed to show the
existence of PBHs depends on the BH mass 
\cite{tisserand,wyrzykowski,belotsky1,niikura,blaineau,mroz}. For instance,
gravitational lensing of background stars is the most
powerful method to search for sub-solar PBHs.
Accretion and dynamical effects on baryonic matter become
more important for heavier PBHs.

	There are various scenarios for primordial black hole production, such as: formation of PBHs 
	during the reheating phase at the
	end of inflation \cite{carr3}, inflaton potential with inflection 
	\cite{germani,khlopov1,garcia-bellido,kannike,prokopec,kamenshchik1,ballesteros,figueroa}, sound speed resonace during inflation \cite{cai,chen2}, null-energy condition violation during inflation \cite{cai2}, first order
	phase transitions \cite{hawking2}, and inflation models with non-canonical kinetic terms \cite{yi,papanikolaou}.

	We aim to investigate PBH production in the early Universe in two inflation 
models  both with a potential with inflection and noncanonical kinetic terms,
in particular the Dirac-Born-Infeld (DBI) type of Lagrangian.
The models in which the kinetic term has a noncanonical form are usually
referred to as $k$-essence \cite{armendariz2} or, in the context of inflation, $k$-inflation models \cite{armendariz1}. 
In this work, we study two $k$-essence models using the Hamiltonian formalism for the background field equations.

First, we investigate a $k$-essence  model proposed by Papanikolaou, Lymperis, Lola, and Saridakis \cite{papanikolaou}. We will refer to their model as the PLLS model.
	The reason why we study this model is twofold: a) the model serves as a good example worth comparison with another model we study in this paper, and b) the results obtained here are modified due to correcting a numerical mistake of Ref.\ \cite{papanikolaou}(see footnote 1 on page 6). 
The PLLS Lagrangian is of the form
\begin{equation}
	\mathcal{L}=	\left(\frac{g^{\mu\nu}\varphi_{,\mu}\varphi_{,\nu}}{2}\right)^\alpha-U(\varphi),
\end{equation} 
where $\varphi$ is a scalar field and  {\rm $\alpha\geq 0$} is a constant.
The same form of $k$-essence has recently been studied in the context of the black-bounce solution \cite{pereira}.
The potential $U$ in Ref.\ \cite{papanikolaou} is assumed to be a positive function of $\varphi$ with a specific inflection point. 
%The parameter $M$ has a dimension of mass. 
An inflationary potential with an inflection or near inflection point leads to an ultra slow-roll regime, which in turn leads to a peak in the power spectrum of curvature perturbations on comoving wavenumber scales above the CMB pivot scale $q_{\rm CMB}$.
As a result, the enhanced cosmological perturbations could collapse into PBHs.

Second, we consider the tachyon inflation model
\cite{fairbairn,frolov,shiu1,sami,shiu2,kofman,cline,salamate2018,barbosa2018,dantas2018,steer,bilicJCAP,kamenshchik2,wang}. 
The first proposals for tachyon inflation models date back to the beginning of this century, emerging as promising candidates for a natural mechanism to trigger inflation. In our previous works \cite{bilicJCAP,Bilic:2013dda, Bilic:2017orf,Bilic:2016fgp, Stojanovic:2023qgm}, we have studied tachyon inflation in the contexts of the Randall\textendash{}Sundrum type braneworld cosmology, and have shown that the predicted observational cosmological parameters are in good agreement with Planck data. In this work, we apply the tachyon inflation model to study the formation of PBHs. 

The Tachyon model is motivated by low-energy string theory.
The existence of tachyons in the perturbative spectrum of string theory shows that the perturbative vacuum is unstable. This instability implies that there exists a true vacuum to which a tachyon field $\varphi$ tends \cite{gibbons}.
The foundations of this process are given by a model of effective field theory \cite{sen}
with a Lagrangian of the DBI form
\begin{equation}
	{\cal{L}} = -U(\varphi)
	\sqrt{1-g^{\mu\nu}\varphi_{,\mu}\varphi_{,\nu}} \, .
		\label{eq1000}
\end{equation}
	This model belongs to a wider class of DBI theories described by the Lagrangian \cite{alishahiha}
	\begin{equation}
		{\cal{L}} = -f(\theta)^{-1}
		\sqrt{1-f(\theta)g^{\mu\nu}\theta_{,\mu}\theta_{,\nu}} -V(\theta)\, .
		\label{eq1001}
	\end{equation}
	Making use of a field redefinition $\varphi=\int f(\theta) d\theta$,
	identifying $f(\theta)^{-1}=U(\varphi)$, and setting $V(\theta)=0$, yields a transformed Lagrangian of the form
	(\ref{eq1000}).
	A  Lagrangian of the form (\ref{eq1001}) appears in the so-called brane inflation models \cite{chen2,shandera}. In these models,  the motion of a D3-brane in a warped throat region of a compact space drives inflation, and the DBI field corresponds to the position of the D3-brane.
	In Ref.\ \cite{chen2}, a specific
	form of the warp factor $f$ is chosen to facilitate  
	the phenomenological
	oscillating sound speed. In contrast, the tachyon potential $U$, corresponding to $f^{-1}$
	in Eq.\ (\ref{eq1001}), is chosen
	to have an inflection point, as in the PLLS case, and thereby yields enhanced PBH production.
	
	The advantage of the $k$-essence approach to PBH formation is due to its simple mechanism of ultra-slow-roll dynamics as a consequence of the inflection point. The models of this type yield straightforward predictions easily confronted with observations. Besides, as we show in the last section (see also Refs.\ \cite{alishahiha}), the amplitude of the primordial non-Gaussianity in both PLLS and Tachyon models is well within observational constraints. More generally, the merit of the tachyon model considered here is in its fundamental origin in low-energy string theory and brane dynamics.

To achieve our primary objective, we first calculate the power spectra of the curvature perturbations in the two models mentioned above and demonstrate that the spectra exhibit a substantial enhancement, depending on the input parameters and initial values of the background fields. We calculate the spectra at the acoustic horizon, assuming the Bunch-Davies initial conditions in the deep subhorizon region. Then, using the critical collapse and peaks formalism,  we calculate the PBH abundance today that constitutes a fraction of dark matter.

The remainder of the paper is organized as follows. In the next two sections, we investigate the background equations of the PLLS model (Sec.\ \ref{plls}) and the Tachyon model (Sec.\ \ref{tachyon}). In Section \ref{perturbations}, we study
the spectra of curvature perturbations in both models. In Section \ref{formation}, we present our results on the PBH formation.
In the last section, Sec.\ \ref{conclude}, we summarize our main results and outline conclusions.
In  Appendix \ref{threshold} we outline the
procedure for calculating the critical overdensity threshold for collapse.

{\bf Notation}
\\
We use the metric signature $(+,-,-,-)$. Unless specified otherwise, we use the system of units in which $c=\hbar=1$.
For convenience, we introduce a length scale $\ell$ so that $\ell\gg 1/M_{\rm Pl}$, where
$M_{\rm Pl}=\sqrt{1/(8\pi G}$ is the reduced Planck mass.
By $\mathcal{L}$ and $\mathcal{H}$, we
denote the Lagrangian and Hamiltonian multiplied by $\ell^4$ so that our $\mathcal{L}$ and $\mathcal{H}$ are dimensionless.
By $\varphi$ and $\phi$, we denote the scalar fields connected by $\varphi=\ell^2\phi$ and have the dimension of length and mass, respectively.

%\subsection{Model A}
\section{The PLLS Model}
\label{plls}
In this section, we briefly recapitulate the model of Ref.\ \cite{papanikolaou}.
The Lagrangian of this model is a sum of a non-canonical kinetic term and a potential term with inflection,
\begin{equation}
	\mathcal{L}=
	\left(\frac{X}{2}\right)^\alpha-U(\varphi),
	\label{eq0035}
\end{equation}
where
\begin{equation}
	X=g^{\mu\nu}\varphi_{,\mu}\varphi_{,\nu}
	\label{eq6001}
\end{equation}
is assumed positive.
Here, $\varphi$ is a scalar field of dimension of length, and $\alpha \geq 1$ is a constant. 
The physical field $\phi$,  related to $\varphi$ via
\begin{equation}
	\phi=\varphi/\ell^2,
	\label{eq4036}
\end{equation}
has the usual dimension of mass. 
The potential with inflection proposed in \cite{papanikolaou} is given by
\begin{equation}
	U=\ell^4 V_0\left[\exp\left(-\lambda\frac{|\varphi_0-\varphi|^n {\text{sgn}}(\varphi_0-\varphi)}{\ell^{2n}M_{\rm Pl}^n}\right)
	-\exp\left(-\lambda\frac{\varphi_0^n}{\ell^{2n}M_{\rm Pl}^n}\right)\right] .
	%	U(\lambda\varphi)=V(\lambda\varphi)-V(0),	
	\label{eq1135}
\end{equation}
The quantity $V_0$ is a constant of the dimension of mass to the 4th power
and $\lambda$ is a positive dimensionless constant.
The field shift $\varphi_0$  corresponds to a shift $\phi_0$ of the physical field $\phi$. 
This potential has an inflection at $\varphi=\varphi_0$.

The background evolution will be studied using the covariant Hamilton formalism
(for details see Ref.\ \cite{bilic}).
	For a general scalar field Lagrangian 
	$\mathcal{L}(\varphi_{,\mu},\varphi)$,
	the covariant Hamiltonian ${\cal H}(\eta^\mu, \varphi)$
	is related to
	${\cal L}$ 
	through the Legendre transformation
	\begin{equation}
		{\cal H} (\eta^\mu, \varphi)= \eta^\mu\varphi_{,\mu} -{\cal L} (\varphi_{,\mu}, \varphi) ,
		\label{eq0024}
	\end{equation}
	with conjugate variables satisfying the conditions
	\begin{equation}
		\varphi_{,\mu} =  \frac{\partial{\cal H}}{\partial\eta^{\mu}},
		\label{eq0025}
	\end{equation}
	\begin{equation}
		\eta^\mu =  \frac{\partial{\cal L}}{\partial\varphi_{,\mu}}.
		\label{eq0026}
	\end{equation}
	By making use of Eqs. (\ref{eq0024}), (\ref{eq0026}),  and the Euler-Lagrange equation 
	\begin{equation}
		\left(\frac{\partial{\cal{L}}}{\partial\varphi_{,\mu}}\right)_{;\mu}
		=\frac{\partial{\cal{L}}}{\partial\varphi} ,
		\label{eq0029}
	\end{equation}
	we find 
	\begin{equation}
		{\eta^\mu}_{;\mu}=-\frac{\partial{\cal H}}{\partial\varphi} .
		\label{eq2026}
	\end{equation}
	Equations (\ref{eq0025}) and (\ref{eq2026}) are the covariant Hamilton equations. 
	The most general $k$-essence Lagrangian is a function of the
	form $\mathcal{L}= \mathcal{L}(X,\varphi)$,
	where $X$ is given by (\ref{eq6001}). 
	%we assume that $X\equiv g^{\mu\nu}\varphi_{,\mu}\varphi_{,\nu}$ is postive.
	From (\ref{eq0026}) it follows that the quantity
	\begin{equation}
		\eta^2=g_{\mu\nu}\eta^\mu\eta^\nu=4X(\mathcal{L}_X)^2 
		\label{eq6012}
	\end{equation}
	is positive.
	Here, the subscript $_X$ denotes a derivative with respect to $X$.
	Then, one can introduce the four-velocity
	\begin{equation}
		u_\mu=\frac{g_{\mu\nu}\eta^\nu}{\eta}= \epsilon\frac{\varphi_{,\mu}}{\sqrt{X}},
		\label{eq0032}
	\end{equation}
	where, $\epsilon$ is $+1$  or $-1$ according to whether $\varphi_{,0}$ is respectively positive
	or negative. This choice of $\epsilon$ guarantees that $u_0$ is always positive.
	The quantity $\eta$ in (\ref{eq0032}) is a square root of (\ref{eq6012}),
	\begin{equation}
		\eta \equiv 2\epsilon\sqrt{X}{\cal L}_X= 
		\epsilon \frac{\partial{\cal L}}{\partial\sqrt{X}}.
		\label{eq6002}
	\end{equation}
	We can regard the variable $\eta$ as a conjugate to $\sqrt{X}$ as described in \cite{bilic}.
	Next, using (\ref{eq0024}) and (\ref{eq0026}) one can express $\mathcal{H}$ as a function of $X$ and $\varphi$,  
	\begin{equation}
		{\cal H} = 2X {\cal L}_X -{\cal L} .
		\label{eq6004}
	\end{equation}
	Then, from (\ref{eq6012}) it follows that 
	$X$ is an implicit function of $\eta^2$ and $\varphi$ which we can 
	(in principle) solve for $X$ to obtain
	$X=X(\eta^2,\varphi)$. Hence, in a $k$-essence type of theory, 
	one can regard the Hamiltonian  as a function of 
	$\eta^2=g_{\mu\nu}\eta^\mu\eta^\nu$ and $\varphi$. 
	%Then, we can plug in this solution
	%into the right-hand side of (\ref{eq6004}) to obtain  $\mathcal{H}= %\mathcal{H}(\eta^2,\varphi)$.
	
	Using the definitions of the velocity (\ref{eq0032}) and expansion rate
	\begin{equation}
		H\equiv\frac{1}{3}{u^\mu}_{;\mu} \,  ,
	\end{equation}
	the covariant Hamilton equations
	(\ref{eq0025}) and (\ref{eq2026}) can be expressed as 
	\begin{equation}
		\dot{\varphi} =  \frac{\partial{\cal H}}{\partial\eta},
		\label{eq0033}
	\end{equation}
	\begin{equation}
		\dot{\eta} +3H \eta = - \frac{\partial{\cal H}}{\partial\varphi}.
		\label{eq3033}
	\end{equation}
	where a quantity $f$ with an over-dot means $\dot{f}\equiv u^\mu f_{,\mu}$. In the comoving frame, the
	over-dot becomes a derivative with respect to time, and the expansion rate becomes $H=\dot{a}/a$ as usual.
For the PLLS Lagrangian (\ref{eq0035}) we find
\begin{equation}
	\mathcal{H}=	
	(2\alpha-1)\left(\frac{X}{2}\right)^{\alpha}+
	U(\varphi).
	\label{eq2035}
\end{equation}
In this model, the variable $X$ can be expressed as an explicit function of $\eta$.
By making use of (\ref{eq0035}) and (\ref{eq6002}) we find
\begin{equation}
	\frac{X}{2}=\left( \frac{\eta^2}{2\alpha^2}\right)^{1/(2\alpha-1)}.
	\label{eq0036}
\end{equation}
Then, we obtain the Hamiltonian expressed in terms of $\varphi$ and $\eta$ as
\begin{equation}
	\mathcal{H}=
	(2\alpha-1)\left( \frac{\eta^2}{2\alpha^2}\right)^{\alpha/(2\alpha-1)}+
	U(\varphi) .
	\label{eq0037}
\end{equation}

As usual, the pressure $p$ and energy density $\rho$ are  identified with the
Lagrangian and the Hamiltonian, respectively.
Then, 
the Hubble expansion rate $H$ is given by
\begin{equation}
	H^2=\frac{\mathcal{H}}{3\ell^4 M_{\rm Pl}^2} ,
	\label{eq3060}
\end{equation}
and the speed of sound by
\begin{equation}
	c_{\rm s}^2 \equiv 
	\frac{\mathcal{L}_X}{\mathcal{H}_X}
	%=\frac{\eta(\pi^{1/(2\alpha-1)})_{,\pi}}{\pi^{1/(2\alpha-1)}}
	=\frac{1}{2\alpha-1}.
	\label{eq3029}
\end{equation}
The first slow roll parameter is  given by
\begin{equation}
	\varepsilon_1\equiv -\dot{H}/H^2=
	\frac{\mathcal{L}+\mathcal{H}}{2\ell^4 M_{\rm Pl}^2H^2}=
	\frac{3\alpha\left[ \eta^2/(2\alpha^2)\right]^{\alpha/(2\alpha-1)}}{(2\alpha-1)\left[ \eta^2/(2\alpha^2)\right]^{\alpha/(2\alpha-1)}+
		U(\varphi)}.
	%	\label{eq0046}
\end{equation}
The higher order slow roll parameters $\varepsilon_i$ are defined recursively
\begin{equation}
	\varepsilon_{i+1}=\frac{\dot{\varepsilon}_i}{H\varepsilon_i}, \quad i\geq 1.
\end{equation}
We adopt the convention that inflation ends when either $\varepsilon_1$ or $\varepsilon_2$ 
is close to unity.

The Hamilton equations (\ref{eq0033}) and (\ref{eq3033})
govern the background dynamics.
%where $H$ is the Hubble rate.
Introducing the  e-fold number 
\begin{equation}
	N=\int dtH ,
	\label{eq3030}
\end{equation}
it is convenient to rewrite the Hamilton equations (\ref{eq0033}) and (\ref{eq3033}) 
as differential equations with respect to $N$. In this model we find
\begin{equation}
	\frac{d\varphi}{dN} =\frac{\eta}{\alpha H}\left( \frac{\eta^2}{2\alpha^2}\right)^{(1-\alpha)/(2\alpha-1)} ,
	\label{eq7036}
\end{equation}
\begin{equation}
	\frac{d\eta}{dN}= -3\eta  - \frac{1}{H}\frac{\partial{U}}{\partial\varphi} ,
	\label{eq7034}
\end{equation}
where 
\begin{equation}
	H=\frac{1}{\sqrt3\,\ell^2 M_{\rm Pl}}\left[	(2\alpha-1)\left(\frac{\eta^2}{2\alpha^2}\right)^{\alpha/(2\alpha-1)}
	+ U(\varphi)
	\right]^{1/2}.	 
	\label{eq7035}
\end{equation}

\subsubsection*{The Klein-Gordon equation}

Instead of the two first-order Hamilton equations, one can use the second-order field equation of motion. This equation can be obtained directly from the Lagrangian or combining Eqs.\ (\ref{eq7036}) and (\ref{eq7034}). 
Either way one finds\footnote{
	A comparison with Ref.\ \cite{papanikolaou} reveals that the factor $\alpha (2\alpha-1)$ in the denominator of the last term on the left-hand side
	of Eq.\ (\ref{eq800}) is missing in their equation (2.9).}
\begin{equation}
	\varphi\,^{\prime\prime}+\left(\frac{3}{2\alpha-1}-
	\varepsilon_1\right) \varphi\,^{\prime}
	+\frac{1}{U}\,\frac{dU}{d\varphi}\,
	\frac{3\alpha-(2\alpha-1)\varepsilon_1}{2\alpha(2\alpha-1)\varepsilon_1}
	\, \varphi\,^{\prime 2}=0,
	\label{eq800}
\end{equation}
where 
\begin{equation}
	\varepsilon_1=\frac{\alpha}{\ell^2H^2}\left(\frac{H^2  \varphi\,^{\prime 2}}{2}\right)^\alpha ,
	\label{eq802}
\end{equation}
and $H$ is a solution to  
\begin{equation}
	\ell^2	H^2=\frac{1}{\ell^2M_{\rm Pl}^2}\left[	(2\alpha-1)\left(\frac{H^2{\varphi'}^2}{2}\right)^\alpha
	+ U(\varphi)
	\right].	 
	\label{eq8035}
\end{equation}
Here and from here on, the prime $^\prime$ denotes a derivative with respect to $N$. 
The expression (\ref{eq8035}) is an algebraic equation for the unknown $H^2$, so
the Klein-Gordon equation (\ref{eq800}) involves an implicit function  $H=H(\varphi,\varphi')$.
 \begin{figure}[ht]
 	\centering
		\includegraphics[width=\textwidth ]{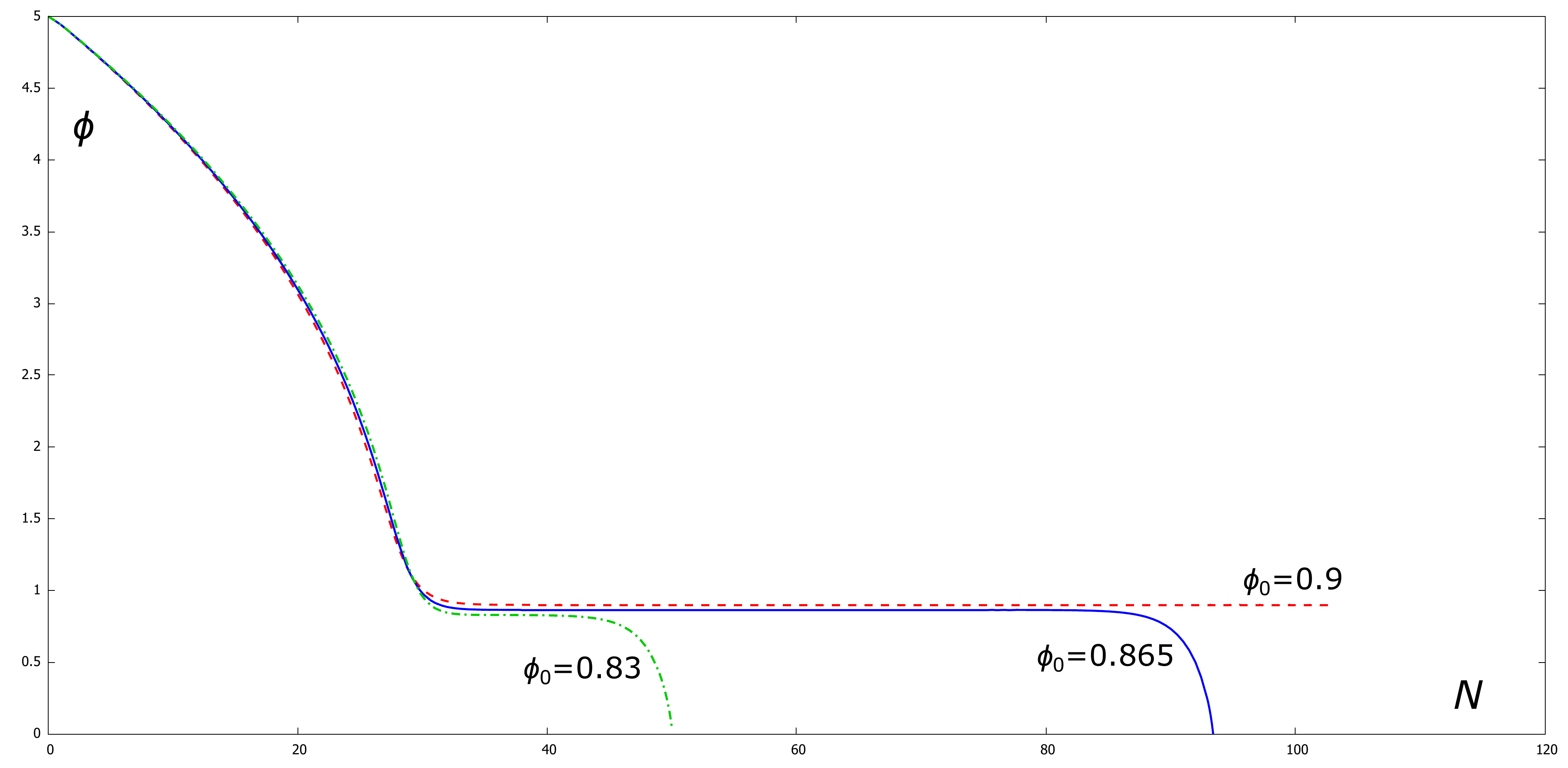}
		\caption{The  solution to the Hamilton equations  
			(\ref{eq7036}) and (\ref{eq7034}) for 
			the PLLS model. The physical field 
			$\phi$ in units of $M_{\rm Pl}$ is ploted
			for    $\alpha=1.5$, fixed initial $\phi_{\rm in}=5 M_{\rm Pl}$, 
			$\phi_{\rm in}'= -8\cdot 10^{-7} M_{\rm Pl}$
			and various $\phi_0$ in units of $M_{\rm Pl}$.
			The remaining parameters $V_0=10^{-16} M_{\rm Pl}^4$ and
			$\lambda=7.54\cdot 10^{-6}$   are as in Ref \cite{papanikolaou}. 
		}
		\label{fig2}
\end{figure}

\subsubsection*{Input parameters and initial values}

The abundance of PBHs depends crucially on the amplitude of the inflationary power spectrum. 
The initial curvature perturbations will eventually collapse to form BHs if the peak in the amplitude exceeds a certain threshold, which strongly depends on the shape of the curvature spectrum \cite{germani}. 
To achieve a desirable spectrum, it is necessary to tune  
the input parameters and initial values.
However, the tuning of these parameters is not quite arbitrary since the normalization of the curvature  spectrum is constrained by its
observational value at the pivot CMB scale.
As a starting point, we can use the parameterization  of Ref.\ \cite{papanikolaou}
and fix $\ell=10^6 M_{\rm Pl}^{-1}$,
$\alpha=1.5$, $V_0=10^{-16}M_{\rm Pl}^4$, and $\lambda=7.54\cdot 10^{-6}$
for $n=3$.
%If $n\neq 3$, to obtain the same value of $U$ at the inflection 
%point, we need to adjust the values of $V_0$ and $\lambda$. 
% In Fig.\ \ref{fig1} we show a comparison
%between potentials with $n=2$ and $n=3$.

Regarding the initial values for $\varphi$ and $\eta$,
we first choose  
$\phi_{\rm in}$ and 
$\phi_{\rm in}'$ in units of $M_{\rm Pl}$ as in Ref.\ \cite{papanikolaou}.
Then, the corresponding initial $\varphi_{\rm in}$ and $\varphi_{\rm in}'$
are obtained using $\varphi=\ell^2\phi$.
The corresponding initial $\eta_{\rm in}$ is obtained by making use of (\ref{eq7036}) and numerically solving   
\begin{equation}
	(\varphi_{\rm in}')^2=\frac{6}{M_{\rm Pl}^2}\,
	\frac{x^{2/(2\alpha-1)}}{(2\alpha-1)x^{2\alpha/(2\alpha-1)}+
		U(\varphi_{\rm in})}	
\end{equation}
for $x$, where 
\begin{equation}
	x=-\frac{\eta_{\rm in}}{\sqrt2 \alpha}.
\end{equation}
Following Ref.\ \cite{papanikolaou},
we tune $\phi_{\rm in}$ and 
$\phi_{\rm in}'$ to obtain a plateau-like behavior
as shown in Fig.\ \ref{fig2}.
These plateau-like solutions are required for a significant PBH production, which we will study in Sec.\ \ref{formation}.

\begin{figure}[ht]
	\centering
		\includegraphics[width=\textwidth]{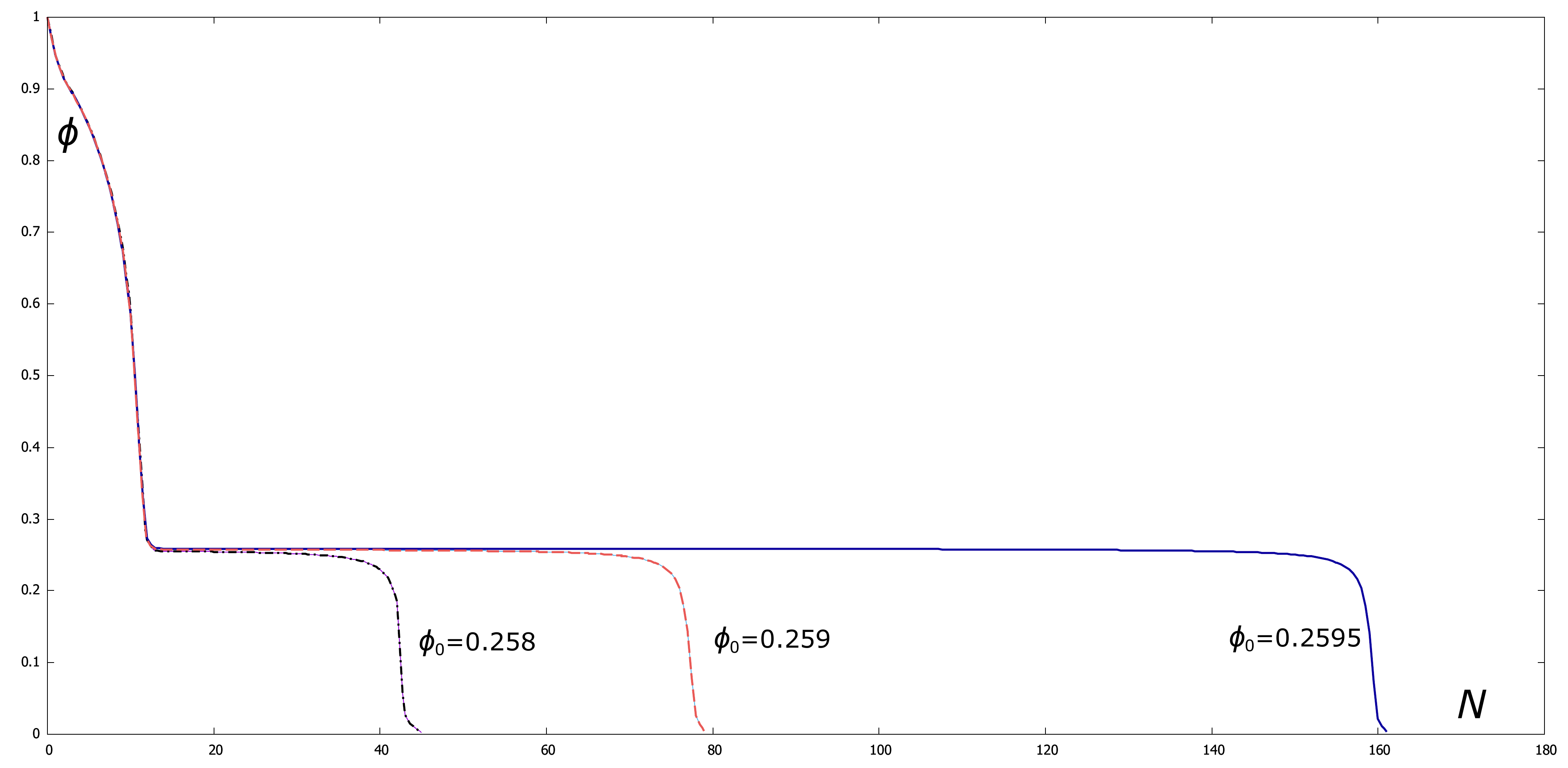}
		\caption{The  solution to the Hamilton equations  
			(\ref{eq3236}) and (\ref{eq3234}) for the Tachyon model. The physical field  $\phi$ 
			(in units of $M_{\rm Pl}$) is plotted as a function of $N$ for $\phi_{\rm in}=1 M_{\rm Pl}$, $ \eta_{\rm in}= -3\cdot 10^3$, $V_0=10^{-16} M_{\rm Pl}^4$,
			$\lambda
			=7.502 \cdot 10^{-6}$, and various field shifts $\phi_0$ in units of $M_{\rm Pl}$, as indicated on the plot.
			%			The remaining imput parameters    are as in Ref \cite{papanikolaou}. 
			%	 $\ell\equiv 1/M$ are as in Ref \cite{papanikolaou}. 
			%The remaining imput parameters $V_0=10^{-16} M_{\rm Pl}^4$ and $\lambda=7.54\cdot 10^{-6}$   are as in Ref \cite{papanikolaou}.  
		}
		\label{fig4}
\end{figure}
\section{The Tachyon model}
\label{tachyon}

The Lagrangian in the Tachyon model is given by
\begin{equation}
	\mathcal{L}=-U(\varphi)\sqrt{1-X},
	\label{eq0038}
\end{equation}
where  
$U$ is a smooth function of the scalar field $\varphi$, and
$X\equiv g^{\mu\nu}\varphi_{,\mu}\varphi_{,\nu}>0$.
As before, we introduce the conjugate field $\eta$
related to $X$ via the definition (\ref{eq6002}). Using (\ref{eq6002}) and (\ref{eq0038})
we obtain $X$ as a function of the fields $\varphi$ 
and $\eta$ 
\begin{equation}
	X=\frac{\eta^2}{U^2+\eta^2}.
	\label{eq0040}
\end{equation}
The Hamiltonian can be expressed as a function of either $X$ or
$\eta^2$,  
\begin{equation}
	\mathcal{H}= \frac{U}{\sqrt{1-X}}
	=\sqrt{U^2+\eta^2}.
	%	\label{eq0041}
\end{equation}
Then, the Hamilton equations (\ref{eq0033})  and (\ref{eq3033}),
expressed as differential equations with respect to $N$, become
\begin{equation}
	\frac{d \varphi}{d  N}=
	\frac{\eta }{H\sqrt{U^2+\eta ^2}} ,
	\label{eq3236}
\end{equation}
\begin{equation}
	\frac{d  \eta }{d  N}= -3\eta   - \frac{U}{H\sqrt{U^2+\eta ^2}}
	\frac{\partial{U}}{\partial\varphi},
	\label{eq3234}
\end{equation}
where the Hubble expansion rate is given by
\begin{equation}
	H^2=
	\frac{\sqrt{U^2+\eta^2}}{3\ell^4 M_{\rm Pl}^2 }.
	%	\label{eq0042}
\end{equation}
The first slow-roll parameter and  the speed of sound squared can be
expressed as 
\begin{equation}
	\varepsilon_1=\frac32	X= \frac32 \frac{\eta^2}{U^2+\eta^2},
	\label{eq1040}
\end{equation}
\begin{equation}
	c_{\rm s}^2 =1-X=\frac{U^2}{U^2+\eta^2}
	=1-\frac23 \varepsilon_1.
	\label{eq3039}
\end{equation}

As in the PLLS model, we choose the potential $U(\varphi)$ defined by (\ref{eq1135}).  
The parameters of the potential are
 $\ell=10^6 M_{\rm Pl}^{-1}$, $n=3$, and
$V_0=10^{-16}M_{\rm Pl}^4$ as before. For this model
we choose a slightly different $\lambda=7.502 \cdot 10^{-6}$.
Regarding the initial values for $\varphi$ and $\eta$,
we first choose  
$\phi_{\rm in}$  
in units of $M_{\rm Pl}$ of the order of those in Ref.\ \cite{papanikolaou}
with the corresponding $\varphi_{\rm in}=\ell^2\phi_{\rm in}$.
 Then, we tune 
$\varphi_{\rm in}$ and $\eta_{\rm in}$ 
to obtain plateau-like solutions as shown in Fig.~\ref{fig4}.

\begin{figure}[t!]
	\begin{center}
%		\includegraphics[width=0.48\textwidth,trim= 0 0cm 0 0cm]{eps1}
%		\hspace{0.02\textwidth}
%		\includegraphics[width=0.48\textwidth]{eps2}
		\includegraphics[width=0.49\textwidth]{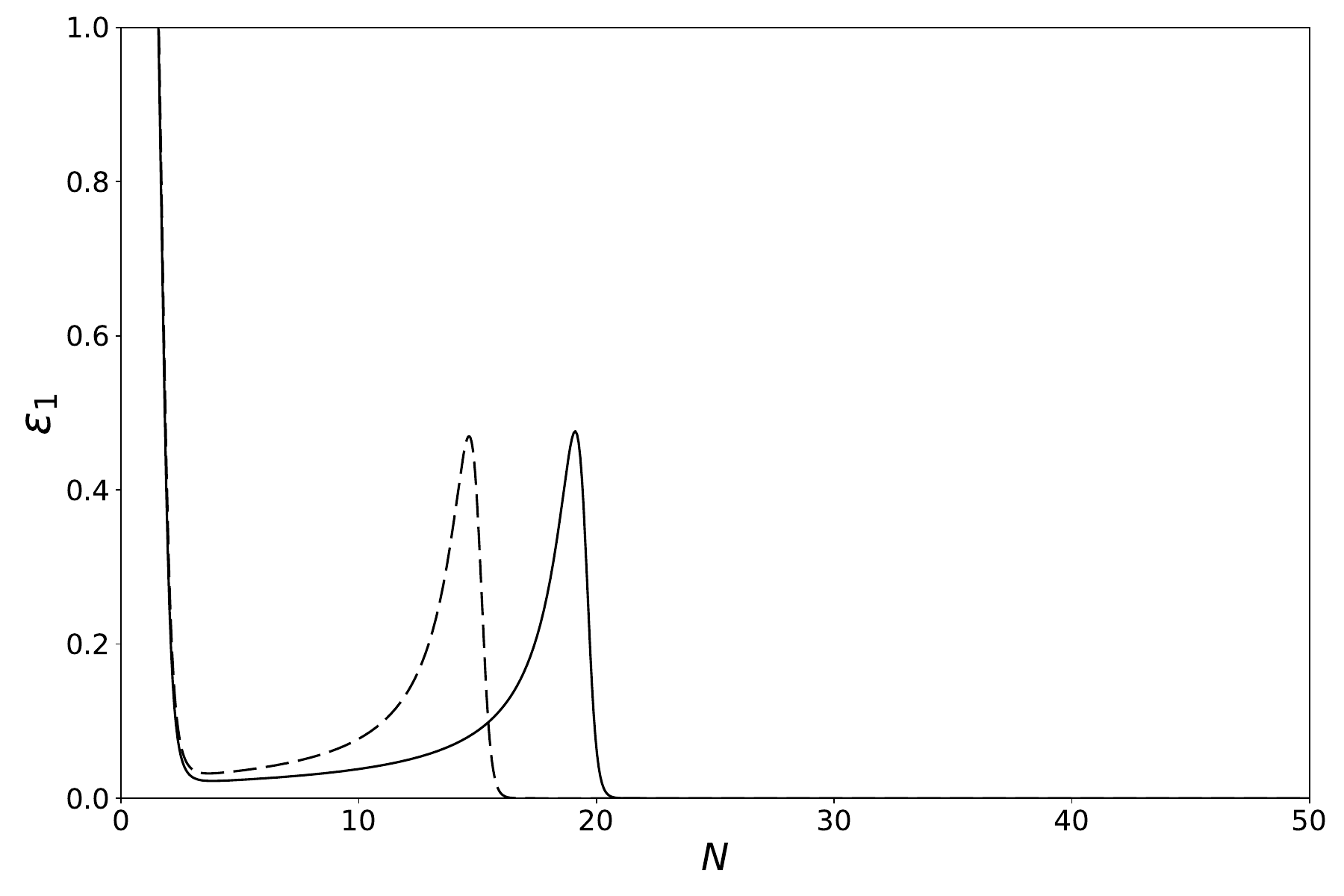}
        \includegraphics[width=0.49\textwidth]{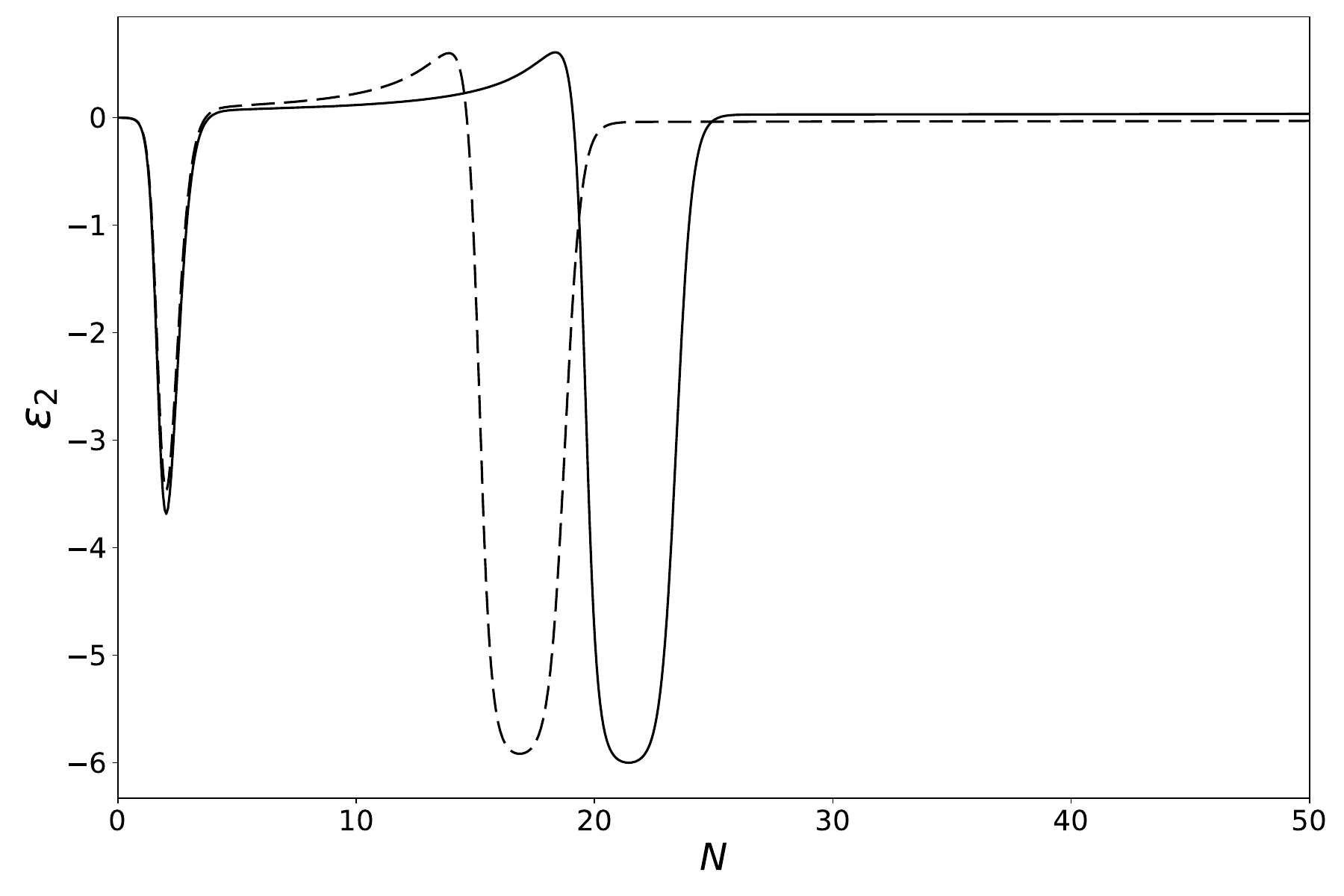}
		\caption{The slow-roll parameters $\varepsilon_1$ (left panel) and $\varepsilon_2$ 
			(right panel) versus $N$ in the Tachyon model  for $\phi_{\rm in}=1.1 M_{\rm Pl}$ (full line) 
			and $1.0520 M_{\rm Pl}$ (dashed line). The remaining corresponding  parameters are listed in Table \ref{table2}.
		}
		\label{fig3}
	\end{center}
\end{figure}

	As in the PLLS case, the scalar field exhibits 
	a plateau owing to the inflection point of the potential at $\phi=\phi_0$. As shown in Fig.~\ref{fig4}, the extension of the plateau depends on $\phi_{0}$. In this flat region, slow-roll conditions will not be met, and inflation will enter a temporary ultra-slow-roll regime. In Fig.\ \ref{fig3} we plot the slow-roll parameters as a function of $N$. 
	During this period, as we will demonstrate in the next section,
	the non-constant mode of the curvature fluctuations, which would decay exponentially 
	in the slow-roll regime, actually grows in the ultra-slow-roll regime. As a consequence, the curvature power spectrum will be enhanced at specific scales that can potentially collapse, forming primordial BHs in the early Universe. 

%
%
%
%
%

%\section{Perturbations}
\section{Curvature perturbations}
\label{perturbations}
We start from  the 1st-order differential equations of Garriga and Mukhanov \cite{garriga} expressed in the form \cite{bilicJCAP,bertini}
\begin{equation}
	a\dot{\xi}_q=z^2c_{\rm s}^2\zeta_q,
	\label{eq3016}
\end{equation}
\begin{equation}
	a\dot{\zeta}_q=-z^{-2}q^2\xi_q,
	\label{eq3017}
\end{equation}
where $q$ is the comoving wavenumber and 
$\zeta_q$ is the Fourier transformed curvature perturbation on uniform-density
hyper-surfaces.
The quantity $z^2$ can be expressed in 
different ways:
\begin{equation}
	z^2=\frac{a^2(p+\rho)}{H^2 c_{\rm s}^2} 
	=\frac{2M_{\rm Pl}^2a^2\varepsilon_1}{c_{\rm s}^2} 	
	=\frac{a^2\eta\mathcal{H}_{,\eta}}{\ell^4H^2 c_{\rm s}^2}	
	=\frac{a^2\mathcal{H}_{,\eta}^2}{\ell^4H^2 \mathcal{H}_{,\eta\eta}} .
	\label{eq4029}
\end{equation}
Here, the subscripts $\eta$ and $\eta\eta$ denote respectively the first and second-order partial derivatives with respect to $\eta$.
%as defined in Appendix \ref{hamilton}, Eq. (\ref{eq4022}). 

It is convenient to substitute the e-fold number $N$ for $t$ using $dt=dN/H$ and
express Eqs.\
(\ref{eq3016})-(\ref{eq3017}) as differential equations with respect to
$N$. Then, combining the obtained equations, 
one can derive 
the Mukhanov-Sasaki second-order differential equation 
\begin{equation}
	\frac{d^2 \zeta_q }{dN^2}+\left(3+\varepsilon_2-\varepsilon_1
	-2\frac{c_{\rm s}'}{c_{\rm s}}\right)\frac{d\zeta_q }{dN}
	+\frac{c_{\rm s}^2 q^2}{a^2 H^2}\zeta_q=0.
	\label{eq7001}
\end{equation}
Equation (\ref{eq7001}) is equivalent to the set (\ref{eq3016})-(\ref{eq3017}) with 
(\ref{eq4029}).
We will
integrate Eq.\ (\ref{eq7001})
in conjunction with the background Hamilton equations for the two models considered in Secs.\ \ref{plls} and 
\ref{tachyon}.

%\subsection{Horizon crossing}
The perturbations travel at the speed of sound, and their horizon is
the acoustic horizon with radius
$c_{\rm s}/H$.
At the acoustic horizon, the perturbations with the comoving wave number $q$ satisfy the horizon crossing relation 
\begin{equation}
	a(N_q) H(N_q)= c_{\rm s}(N_q)q .
	\label{eq50}
\end{equation}
In the slow-roll regime, the perturbations are conserved once they cross the horizon from the subhorizon 
to the superhorizon region. A rough estimate \cite{bilic} shows that
the horizon crossing happens at a relatively large 
$q$-dependent $N$
of the order $N_q\sim 7+\ln q/q_{\rm CMB}$, where $q_{\rm CMB}$ is the CMB pivot scale.
The curvature perturbations $\zeta_q$ are approximately constant at large $N$ once they cross the horizon 
and enter the superhorizon region where $a(N) H(N)> c_{\rm s}(N)q$, $N>N_q$.

%The power spectrum plots will correspond to the solutions to 
 The power spectrum of curvature perturbations is obtained from the two-point correlation function
\begin{equation}
	\langle\hat{\zeta}_q\hat{\zeta}_{q'}\rangle=
	(2\pi)^3 \delta(\mbox{\boldmath $q$}+\mbox{\boldmath $q$}')|\zeta_q|^2,
	\label{eq0063}
\end{equation} 
where $\hat{\zeta}_q$ is the operator associated with the curvature perturbation $\zeta$
\cite{mukhanov}. 
We define 
\begin{equation}
	\mathcal{P}_{\rm S}(q)=\frac{q^3}{2\pi^2}|\zeta_q(N_q)|^2 ,
	\label{eq4024}
\end{equation}
where $\zeta_q(N_q)$ is the solution to Eq.\ (\ref{eq7001})
taken at the point
$N_q$ which for given $q$ satisfies (\ref{eq50}).
%\subsection{Normalization of the spectrum}
The curvature perturbation spectrum  needs to satisfy  
the Harrison-Zeldovich spectrum near $q=q_{\rm CMB}=0.05$ Mpc$^{-1}$
\begin{equation}
	\mathcal{P}_{\rm S}(q)= A_{\rm S} \left(\frac{q}{q_{\rm CMB}}\right)^{n_{\rm S}-1},
	\label{eq4025}
\end{equation}
where $n_{\rm S}$ is the scalar spectral index and $A_{\rm S}=(2.10\pm 0.03)\times 10^{-9}$.
Since the Mukhanov-Sasaki equation (\ref{eq7001}) is linear,
the requirement 
\begin{equation}
	\mathcal{P}_{\rm S}(q_{\rm CMB})=\frac{q_{\rm CMB}^3}{2\pi^2}|\zeta_{q_{\rm CMB}}(N_{q_{\rm CMB}})|^2=A_{\rm S}
	\label{eq5024}
\end{equation}
can fix the normalization of $\zeta_q$.

For convenience, instead of $\zeta_q$ from now on, we will use 
the Mukhanov-Sasaki function $v_q=z\zeta_q$, 
where the quantity $z$ is defined by (\ref{eq4029}).
Then,  Eq.\ (\ref{eq7001}) may be written as
\begin{equation}
%\begin{eqnarray}
	v''_q+(1-\varepsilon_{1})v'_q+\left[
	\frac{c_{\rm s}^2q^2}{a^2H^2}	
	-(2-\varepsilon_{1}+\frac{\varepsilon_2}{2}-\frac{c_{\rm s}'}{c_{\rm s}})(1+\frac{\varepsilon_2}{2}-\frac{c_{\rm s}'}{c_{\rm s}})
	%	\right.
	%	\nonumber \\
	%	\left.		
	-\frac{\varepsilon_{2}\varepsilon_3}{2}+(\frac{c_{\rm s}'}{c_{\rm s}})'\right]v_q=0.
	\label{eq7002}
%\end{eqnarray}
\end{equation}
Note that in the PLLS model $c_{\rm s}'$ vanishes whereas in the Tachyon model, we have
\begin{equation}
	\frac{c_{\rm s}'}{c_{\rm s}}=-\frac{\varepsilon_{1}\varepsilon_{2}}{3-2\varepsilon_{1}},
\label{eq701}
\end{equation}
\begin{equation}
	(\frac{c_{\rm s}'}{c_{\rm s}})'=-\frac{(\varepsilon_{1}\varepsilon_{2}^2+\varepsilon_{1}\varepsilon_{2}\varepsilon_{3})
		(3-2\varepsilon_{1})+2\varepsilon_{1}^2\varepsilon_{2}^2}{(3-2\varepsilon_{1})^2}.
	\label{eq702}
\end{equation} 
Note that Eq.\ (\ref{eq7002}) follows from Eqs.\ (\ref{eq3016})-(\ref{eq3017}) via (\ref{eq4029}) and (\ref{eq7001}). Since the derivation of Eqs.\ (\ref{eq3016})-(\ref{eq3017}) does not involve the slow-roll assumption, the quantities $\varepsilon_1$ and $\varepsilon_2$ that appeear in Eqs.\ (\ref{eq7002})-(\ref{eq702}) are not necessarily small.
In our numerical integration of Eq.\ (\ref{eq7002}) with (\ref{eq701})-(\ref{eq701}),
we will treat the slow-roll parameters $\varepsilon_1$ and $\varepsilon_2$ as exact background quantities, not subject to the slow-roll conditions.
%Consequently, in our further numerical simulations, we will not assume the slow-roll conditions.

To solve Eq.\ (\ref{eq7002}), we need to choose the initial point $N_{{\rm in},q}$ for each wavenumber $q$. The appropriate $q$-dependent initial point must be in the deep subhorizon region where 
$a_{\rm in}H_{\rm in}\ll c_{\rm s}q$.
More precisely, for each wave number $q$ that satisfies the horizon crossing relation
$a(N_q)H(N_q)=c_{\rm s}(N_q) q$,
we start at a $q$-dependent $N_{{\rm in},q} <N_q$ such that
\begin{equation}
	a(N_{{\rm in},q})H(N_{{\rm in},q})=\beta c_{\rm s}(N_{{\rm in},q})q ,
	\label{eq20}
\end{equation}
where $\beta$ is a small parameter, e.g., $\beta=0.01$ as proposed in Ref.\ \cite{de}.
Owing to the $N$-translation invariance of the background equations,
we could choose an arbitrary origin of inflation $N_0$ as an initial point related to
a wavenumber  
$q_0< q_{\rm CMB}$. In other words, we set
\begin{equation}
	a_0	H_0=\beta c_{\rm s0}q_0=\epsilon c_{\rm s0}q_{\rm CMB} ,
	\label{eq32}
\end{equation}
where $a_0=a(N_0)$,
$H_0= H(N_0)$, $c_{\rm s0}= c_{\rm s}(N_0)$,
and the small parameter 
$\epsilon$  satisfies  $\epsilon< \beta$, e.g., $\epsilon=0.001$.
Then, using $a=a_0 e^{N-N_0}$ together with  
(\ref{eq32}), we can write the first term in the square brackets in Eq. (\ref{eq7002})
as 
\begin{equation}
	\frac{c_{\rm s}^2q^2}{a^2H^2}=
	\frac{e^{2N_0-2N}}{\epsilon^2} \frac{c_{\rm s}^2}{c_{{\rm s}0}^2}\frac{H_0^2}{H^2} \frac{q^2}{q_{\rm CMB}^2}.
	\label{eq43}
\end{equation}
For more details on the choice of the initial point, see  Ref.\ \cite{bilic}.

To find the spectrum at the horizon crossing, equation (\ref{eq7002}) should  be integrated for a fixed $q$ up to the e-fold number
$N_q$, which satisfies the horizon-crossing equation 
(\ref{eq50}). Fixing the beginning of inflation at $N_0=0$ and combining Eqs.\ (\ref{eq50}) and  
(\ref{eq32}) we obtain
\begin{equation}
	e^{N_q}= \frac{1}{\epsilon} \frac{q}{q_{\rm CMB}} 
	\frac{H_0}{H(N_q)}\frac{c_{\rm s}(N_q)}{c_{{\rm s}0}},
	\label{eq7006}
\end{equation}
which, given $q$, may be regarded as an algebraic equation for $N_q$. 
The explicit 
functional dependence $H(N_q)$ and $c_{\rm s}(N_q)$
in (\ref{eq7006}) is obtained by integrating the background Hamilton equations
in parallel with (\ref{eq7002}).
Then, the spectrum $\mathcal{P}_{\rm S}(q)$ can be plotted using the solutions $\zeta_q(N_q)$ at the point $N_q$ that satisfies (\ref{eq7006})
for each $q$.

\subsubsection*{Initial conditions}

%in units in which $M_{\rm Pl}=1$.
To determine the proper initial conditions in the deep subhorizon region, we adopt the standard Bunch-Davies vacuum solution \cite{bunch}  
\begin{equation}
	v_q(\tau)=\frac{e^{-ic_{\rm s}q\tau}}{\sqrt{2c_{\rm s}q}} ,
	\label{eq1}
\end{equation} 
where the conformal time $\tau$ is defined by
\begin{equation}
	d\tau= \frac{dt}{a}= \frac{dN}{aH} .
%	\tau=-\frac{1+\varepsilon_1}{aH}+\mathcal{O}(\varepsilon_1^2) ,
	\label{eq2}
\end{equation}
%satisfies $-qc_{\rm s}\tau \gg 1$ in the deep subhorizon region. 
Hence, the initial values of $v_q$ and $v_q'$ at $N_{{\rm in},q}$  are determined 
by (\ref{eq1}) up to an arbitrary phase. The simplest choice is 
\begin{equation}
	v_{q\rm in}\equiv v_q(N_{{\rm in},q})=\frac{1}{\sqrt{2c_{\rm s}q}},
	\label{eq10}
\end{equation}
\begin{equation}
	v'_{q\rm in}\equiv v_q'(N_{{\rm in},q})=-i\frac{c_{\rm s}q}{a(N_{{\rm in},q})H(N_{{\rm in},q})\sqrt{2c_{\rm s}q}}.
	\label{eq11}
\end{equation}
Then, using (\ref{eq20}), we have
\begin{equation}
	v'_{q\rm in}=-i\frac{1}{\beta\sqrt{2c_{\rm s}q}},
	\label{eq13}
\end{equation}
where $\beta$ is a small constant discussed above, e.g., $\beta=0.01$.
It is understood that the function $c_{\rm s}$ 
in Eqs.\ (\ref{eq10})-(\ref{eq13}) 
is taken at $N=N_{{\rm in},q}$.

\subsection{Redefinition of the input parameters and fields}
\label{renormalize}

In our calculations, we assume certain values of the parameters and initial values of the fields. Besides, a solution to Eq.\ (\ref{eq7002}) involves
the Bunch-Davies initial conditions (\ref{eq10}) and (\ref{eq11}). These initial conditions determine the normalization of 
$\zeta_q(N)$ which in turn fixes the normalization of $\mathcal{P}_{\rm S}$. 
Hence,
there is {\it a priory} no guarantee that the obtained spectrum will satisfy 
the condition (\ref{eq5024}). Instead, the obtained spectrum   $\mathcal{P}_{\rm S}$ will satisfy
\begin{equation}
	\mathcal{P}_{\rm S}(q_{\rm CMB})=\frac{A_s}{c_0},
	\label{eq65}
\end{equation}
where $c_0$ is a constant, generally $c_0\neq 1$.
Thus, we have a conflict between the imposed Bunch-Davies vacuum and 
the normalization of the spectrum to the observed value at the CMB pivot scale.  
\begin{table}[h]
	\caption{Input parameters and fields}
	\centering
	\begin{tabular}{||c|c||c|c||} 
		\multicolumn{4}{c}{}  \\	
		\hline
		\multicolumn{2}{||c||}{\textbf{PLLS model}, $\alpha=1.5$} & \multicolumn{2}{|c||}{\textbf{Tachyon model}}  \\ \hline
		original  &redefined  &   original  &redefined  \\ \hline  
		$U$&$c_0^{-1} 	U$ &$U$&$c_0^{-1} 	U$ \\ \hline
		$\lambda$& $ c_0^{(1-\alpha)/(2\alpha)} \lambda$&$\lambda$ &$c_0^{-1/2}\lambda$\\ \hline
		$\varphi$&$c_0^{(\alpha-1)/(2\alpha) } \varphi $ & $\varphi$&$c_{0}^{1/2}\varphi$ \\ \hline
		$\eta$& $c_{0}^{(1-2\alpha)/(2\alpha)}\eta$& $\eta$ &$c_{0}^{-1}\eta$ \\ \hline
	\end{tabular}
	\label{table2}
\end{table}

To rectify this conflict, 
we need to properly redefine the input parameters and initial values.
Using an arbitrary constant $c_0$, one may easily show \cite{bilic} that
the Hamilton equations and the corresponding solutions are invariant under the simultaneous rescaling of the parameters and initial values as shown in Table \ref{table2}.
In both models, the Hubble rate scales according to
$H\rightarrow c_0^{-1/2} H$ 
%with no dependence on $c_0$.
and the rescaled Hamilton equations retain their original form
with no dependence on $c_0$.

\begin{figure}%[h]
	\centering
		\includegraphics[width=\textwidth]{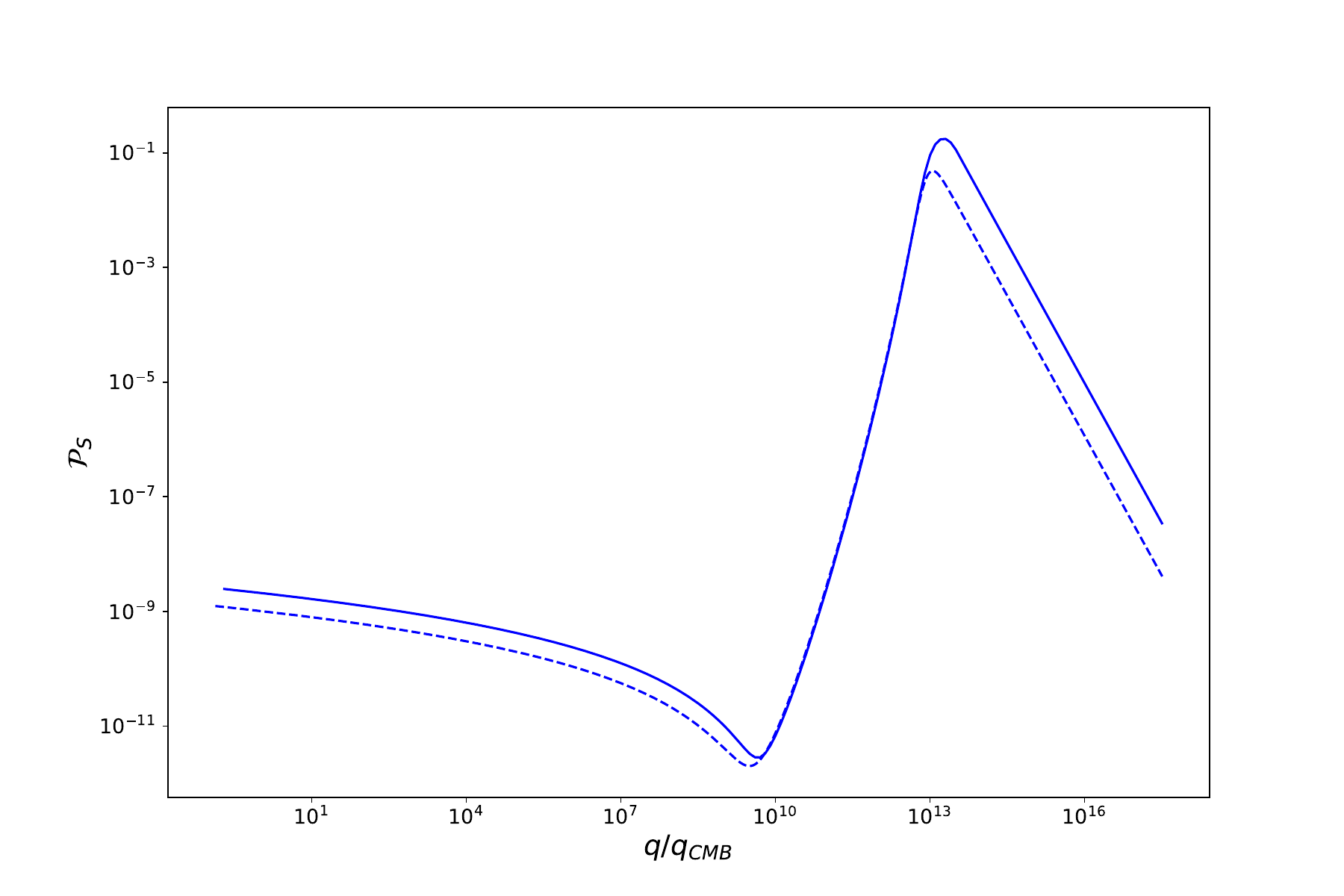}
		\caption{The curvature power spectrum obtained by
			numerically solving Eq.\ (\ref{eq7002}) 
			(full line) and the  spectrum  approximated by Eq.\
			(\ref{eq5021}) (dashed line) for the PLLS inflation model 
			with $n=3$, $\phi_0=0.8354 M_{\rm Pl}$, and 
			$\alpha=1.5$. The values of $V_0$ 		 
			and $\lambda$   are as in Fig.\ \ref{fig2}.   	
			The initial values are 
			$\phi_{\rm in}=5.20473 M_{\rm Pl}$,
			$\eta_{\rm in}=-2.81744\cdot 10^{-7}$. 
			The initial perturbation $v_{q\rm in}$ is determined by the Bunch-Davies vacuum and  $\mathcal{P}_{\rm S}(q_{\rm CMB})$
			as discussed in the text.	
		}
		\label{fig7}
\end{figure}
\begin{figure}%[ht]
	\centering
	\includegraphics[width=\textwidth]{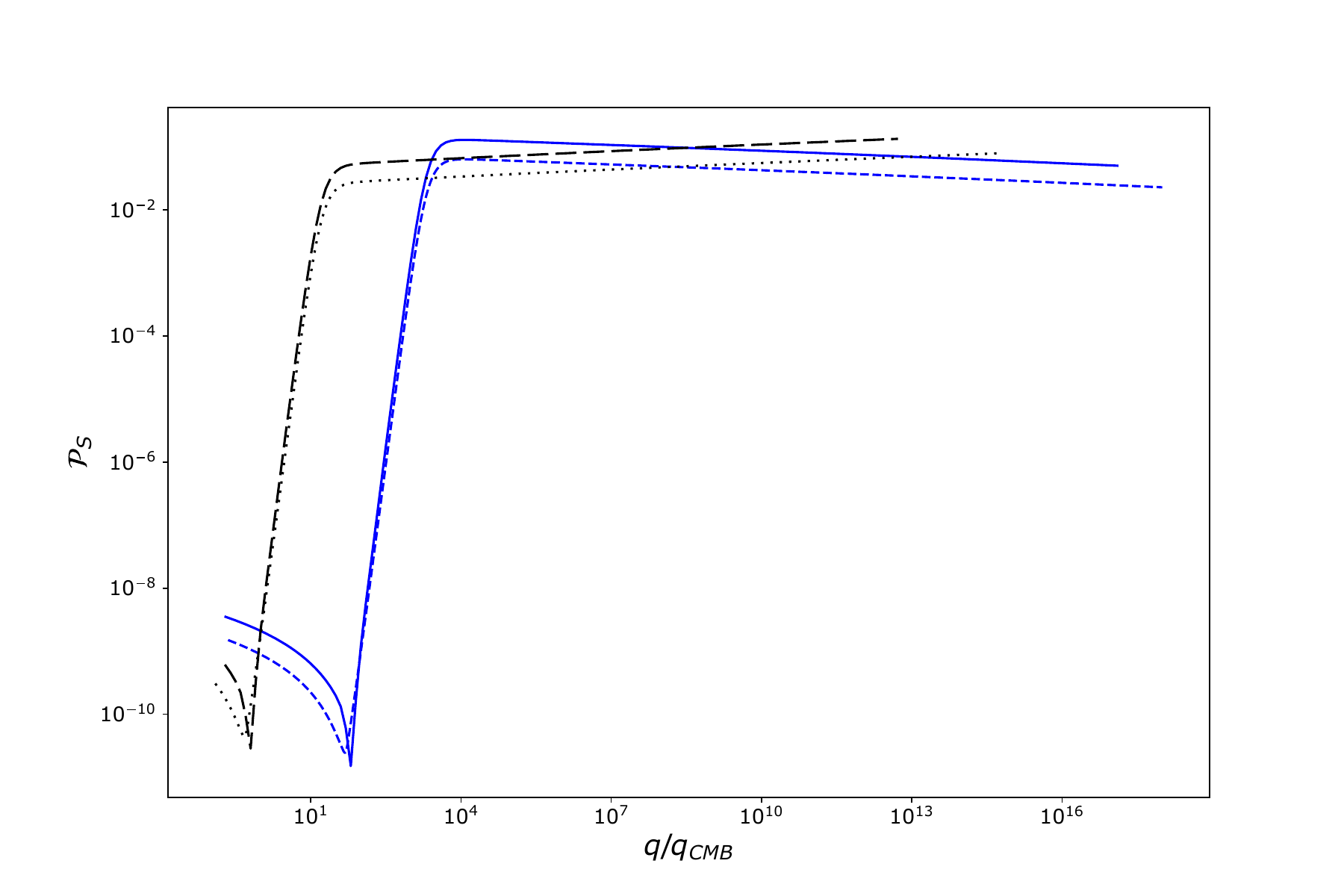}
	\caption{
		The curvature power spectrum obtained by
		numerically solving equation (\ref{eq7002}) 
		(full blue and long-dashed black lines) combined with the  spectrum  approximated by 
		Eq.\ (\ref{eq5021}) (short dashed blue and dotted black lines) for the Tachyon model inflation with the potential
		$U$ as in the PLLS model. The parameters and initial values of the corresponding background solutions are
		$\phi_{0}=0.2595 M_{\rm Pl}$, $V_0=10^{-16} M_{\rm Pl}^4$, $\lambda=7.502 \cdot 10^{-6}$, 
		$\phi_{\rm in}=1.1 M_{\rm Pl}$, and $\eta _{\rm in}=-5.3\cdot 10^4$ (full and short-dashed  lines), 
		and $\phi_{0}=0.259831 M_{\rm Pl}$, $V_0=10^{-16} M_{\rm Pl}^4$, $\lambda
		=7.600 \cdot 10^{-6}$, $\phi_{\rm in}=1.052 M_{\rm Pl}$ and $\eta _{\rm in}=-5.3625\cdot 10^4$
		(long-dashed and dotted lines).}
	\label{fig9} 
\end{figure}

We will use this scaling property of our $k$-essence models to eliminate the constant $c_0$ in Eq.\ (\ref{eq65}) by redefining the input parameters.  To begin, 
we multiply the right-hand side of (\ref{eq4024})  by $c_0$ and write the power spectrum 
$\mathcal{P}_{\rm S}$ as 
\begin{equation}
	\mathcal{P}_{\rm S}=c_0\frac{q^3}{2\pi^2}|\zeta_q(N_q)|^2 = 
	\frac{c_0 H^2}{4\pi^2\varepsilon_1 M_{\rm Pl}^2}
	q|v_q|^2,
	\label{eq66}
\end{equation}
where we have used Eqs.\ (\ref{eq4029}) and  (\ref{eq50}).
Now we absorb $c_0$ in $H^2$ 
and rescale the Hubble rate as $H\rightarrow c_0^{-1/2}  H$.
This rescaling does not affect Eq.\ (\ref{eq7002}) and its solution $v_q$.
Thus, we obtain a properly normalized spectrum without the $c_0$ factor. However, a rescaling of $H$ implies a redefinition of the model input parameters and initial values of the background fields. 
Using  the rescaling defined in Table \ref{table2} 
we infer a redefinition of the parameters.

In this way, if we repeat the calculations using the redefined input parameters
listed in Table \ref{table2}
for both models, we will obtain the power spectra that agree with the observed value at the CMB pivot scale.
 
\subsection{Approximate spectrum}
\label{approximate}

In the slow-roll regime, the curvature spectrum can be approximated by
(see, e.g., Ref.\ \cite{bertini})
\begin{equation}
	\mathcal{P}_{\rm S}(q)\simeq \frac{1}{8\pi^2c_{\rm s}\varepsilon_1}
	\frac{H^2}{M_{\rm Pl}^2} ,
	\label{eq6021}
\end{equation}
where for each $q$, the quantities $H$, $ c_{\rm s}$, and $\varepsilon_1$ take on their horizon crossing values at the corresponding $N_q$.
It is important to stress that in the models considered here the perturbations
evolve through a phase which departs from the slow-roll regime, as is evident from Fig.\ \ref{fig3}. Nevertheless, we will use (\ref{eq6021}) for the sake of comparison. However, we expect a substantial departure of the power spectrum (\ref{eq6021}) from the exact one obtained by numerically integrating Eq.\ (53).

The approximate spectrum obtained in this way will not in general
have the correct observed value at the pivot scale $q_{\rm CMB}$.
To satisfy the proper normalization of $\mathcal{P}_{\rm S}(q)$ at $q=q_{\rm CMB}$,  we can 
introduce a constant factor $\bar{c}_0$. To wit, we define
\begin{equation}
	\mathcal{P}_{\rm S}(q)\simeq \frac{\bar{c}_0}{8\pi^2c_{\rm s}\varepsilon_1}
	\frac{H^2}{M_{\rm Pl}^2}
	\label{eq5021}
\end{equation}
and fix $\bar{c}_0$ so that $\mathcal{P}_{\rm S}(q_{\rm CMB})$ has the correct 
observed value at the pivot scale.
However, a choice  $\bar{c}_0\neq 1$ is not compatible with the assumed Bunch-Davies asymptotic behavior. 
As before, this problem is resolved by absorbing $\bar{c}_0$ in $H^2$ and redefining the physical parameters as shown in Table \ref{table2}.

{\bf Nota Bene}:
The constants $c_0$ and $\bar{c}_0$ need not be necessarily equal.
However,  to compare the exact and approximate spectra, we must stick to the same parameterization of the model in both exact and approximate calculations. In this case, we will replace $\bar{c}_0$ in (\ref{eq5021}) by $c_0$ so that the redefinition of parameters and initial values will be the same.
Obviously, in this case, the approximate spectrum will fail to reproduce the correct
observed value at $q=q_{\rm CMB}$.

We plot the calculated spectrum for the PLLS model in Fig.\ \ref{fig7}.
In  Fig.\ \ref{fig9}, we plot the spectra calculated for the Tachyon model for two sets of parameters. The approximation (\ref{eq6021}) correctly reproduces the shape of the spectrum but underestimates the peak by a factor of the order of $10$ and violates the normalization condition (\ref{eq5024}) (see the note after Eq.\ (\ref{eq5021})).

\section{PBH formation}
\label{formation}

In the radiation-dominated epoch, PBHs could  form
by small-scale cosmological perturbations with sufficiently large overdensities  collapsing
after re-entering the cosmological horizon. Assuming
spherical symmetry, such regions can be described by
the following approximate form of the metric at superhorizon
scales \cite{musco}
\begin{equation}
	ds^2=dt^2-a(t)^2 e^{2\zeta(r)}(dr^2+r^2d\Omega^2) .
\end{equation}
The quantity  $\zeta(r)$ is spherically symmetric comoving curvature perturbation
conserved on superhorizon scales \cite{wands}. The corresponding
density contrast $\delta_r\equiv \delta\rho/\rho(r,t)$ is given by \cite{germani}
\begin{equation}
	\delta_r=-\frac89 \frac{1}{a^2H^2}e^{-5\zeta(r)/2}\nabla^2e^{\zeta(r)/2},
\label{eq3002}
\end{equation}
 Keeping the linear terms only, we find
	\begin{equation}
		\delta_r
		\simeq -\frac49 \frac{1}{a^2H^2}\nabla^2\zeta(r).
		\label{eq3003}
	\end{equation}
	In the momentum space we obtain 
	\begin{equation}
		\delta_q
		\simeq \frac49 \frac{q^2}{a^2H^2}\zeta_q .
		\label{eq001}
	\end{equation}
	Hence, the two-point correlation function (\ref{eq0063}) dominates the power spectrum of the density contrast via the linear relation (\ref{eq001}). We find
	\begin{equation}
		P_\delta\equiv \frac{q^3}{2\pi^2}\left|\delta_q\right|^2
		=\frac{q^3}{2\pi^2}\left(\frac49\right)^2 \frac{q^4}{a^4H^4}\left|\zeta_q\right|^2
		=\left(\frac49\right)^2 \frac{q^4}{a^4H^4}\mathcal{P}_{\rm S} .
	\end{equation}
	In our analysis, we assume that both the density contrast $\delta$ and curvature perturbation $\zeta$ are approximately
	Gaussian variables. In principle, a non-Gaussian contribution to the statistics of the primordial curvature perturbations could modify the PBH abundance. We will provide a brief comment on non-Gaussianity at the end of this section.
	
	As we will shortly see, the PBH mass fraction $\beta(M)$ will be mainly described  by   
	the smoothed variance $\sigma$ of the density contrast $\delta$, with rough functional dependence as
	\begin{equation}
		\beta\propto \left(-\frac{\delta_{\rm c}}{2\sigma^2}\right) ,
	\end{equation}
	where $\delta_{\rm c}$ is the critical overdensity collapse threshold.
	Hence, the larger the variance $\sigma$, the larger the mass fraction $\beta$.
	We define the smoothed variance of $\delta$
	over the smoothing scale $R=1/(aH)$
\begin{equation}
	\sigma(R)^2=\int_0^\infty \frac{dq}{q}	 W(q,R)^2P_\delta
	=\left(\frac49\right)^2\int_0^\infty \frac{dq}{q}(Rq)^4 W(q,R)^2
	\mathcal{P}_{\rm S}
	\label{eq0012}
\end{equation}
and the first moment of the smoothed power spectrum
\begin{equation}
	\mu(R)^2=\int_0^\infty \frac{dq}{q} (Rq)^2	 W(q,R)^2 P_\delta
	=\left(\frac49\right)^2\int_0^\infty \frac{dq}{q}(Rq)^6W(q,R)^2\mathcal{P}_{\rm S},
	\label{eq0013}	
\end{equation}
where $W$ is a smoothing window function.
	It is not
	obvious what the best choice for the smoothing function is. 
	There is a priori no reason to choose one smoothing over another \cite{young3,young5}.
	A common convenient choice is a Gaussian window function, although
	a top-hat smoothing function was occasionally used in the
	past \cite{green,scoccimarro}. However, the top-hat window function is sensitive to scales well within the
	horizon \cite{green2}
	which requires careful treatment \cite{bringmann,blais}.
	
Ando, Inomata, and Kawasaki \cite{ando}
investigated the dependence of the variance $\sigma(R)$ on the choice between $x$-space top hat, gaussian, and  $k$-space top hat window function (see also Refs.\ \cite{ando2,young4}).
In particular, for a simple scale invariant power spectrum $\mathcal{P}_{\rm S}=A_{\rm S}$, they found that the variances $\sigma(R)$ differ in proportions 1:0.29:0.22 for 
the $x$-space top hat, Gaussian, and  $k$-space top hat window functions, respectively.
We have also checked the window function dependence of $\sigma(R)$ assuming a gaussian shape power spectrum $\mathcal{P}_{\rm S}(q) \propto \exp [-R^2 (q-R_0^{-1})^2/2]$ peaked at various $R_0^{-1}$.  For example, for $R_0=R$, we have found the proportions 1:0.36:0.27  for the mentioned window functions, respectively. These proportions do not change significantly with varying $R_0$ in the interval 0.3 - 10, where they remain roughly 1:1/3:1/4.

	Following Green et al.\ \cite{green2}, we choose  gaussian 
\begin{equation}
	W(q,R)=e^{-q^2R^2/2},
\end{equation}
as a more conservative choice than the $x$-space top-hat window function.

%\subsection{The approximate variance}

\subsubsection*{Toy model}

For an illustration, it is useful to consider a 
flat power spectrum parameterized as
\begin{equation}
	\mathcal{P}_{\rm S} \simeq \mathcal{P}_0 \Theta(q-q_{\rm min})\Theta(q_{\rm max}-q) .
	\label{eq0014}
\end{equation}
This power spectrum, which was studied previously in the literature as a toy model (see, e.g.,
\cite{deluca}), 
can serve as a rough approximation to the Tachyon model power spectrum of
Fig.\ \ref{fig9} with $\mathcal{P}_0\simeq 0.1$ - $0.2$, $q_{\rm min}=10^4 q_{\rm CMB}$,  and $q_{\rm max}=10^{16} q_{\rm CMB}$.

Using this in (\ref{eq0012}) and (\ref{eq0013})
we obtain
\begin{equation}
	\sigma^2=\frac12 \left(\frac49\right)^2\mathcal{P}_0\left[\left(1+q_{\rm min}^2R^2\right)e^{-q_{\rm min}^2R^2}
	-\left(1+q_{\rm max}^2R^2\right)e^{-q_{\rm max}^2R^2}	
	\right
	],
	\label{eq0015}
\end{equation}
\begin{equation}
%\begin{eqnarray}
	\mu^2= \frac12\left(\frac49\right)^2\mathcal{P}_0\left[\left(2+2q_{\rm min}^2R^2+q_{\rm min}^4R^4\right)e^{-q_{\rm min}^2R^2}
%	\right.\nonumber\\
%	\left. 
	-\left(2+2q_{\rm max}^2R^2+q_{\rm max}^4R^4\right)e^{-q_{\rm max}^2R^2}	
	\right].
%\end{eqnarray}
\end{equation}
The functions $\sigma(R)$ and $\mu(R)$ have a top-hat shape  with plateau heights
$\sigma_{\rm \max} \simeq (4/9)\sqrt{\mathcal{P}_0/2}$
and $\mu_{\rm \max} \simeq (4/9)\sqrt{\mathcal{P}_0}$.
For $R$ satisfying $1/q_{\rm max}\ll R \ll 1/q_{\rm min}$ we have
\begin{equation}
	\sigma = \frac49\sqrt{\frac{\mathcal{P}_0}{2}}\left(1+q_{\rm min}^2R^2 + \mathcal{O}(q_{\rm min}^4R^4)\right),
\end{equation}
and hence, the maximal value of sigma is
\begin{equation}
	\sigma_{\rm \max} \simeq \frac49\sqrt{\frac{\mathcal{P}_0}{2}} .
	\label{eq0018}
\end{equation}
In this range of $R$, we also have
\begin{equation}
	\frac{\mu}{\sigma}\simeq \sqrt2 + \mathcal{O}(q_{\rm min}^4R^4).
	\label{eq0007}
\end{equation}

\subsubsection*{Functional dependence  $R(M)$} 
Through (\ref{eq0012}), the variance $\sigma$ is a function of 
the comoving horizon radius at the PBH formation time
$R = (a_{\rm f} H_{\rm f})^{-1}=1/q_{\rm f}$, 
where the subscript f refers to the  PBH formation time.
Since the PBH mass $M$ at the formation time is of the order of the cosmological horizon mass, i.e., 
\begin{equation}
	M \simeq M_{H_{\rm f}}=\frac{4\pi}{3}H_{\rm f}^{-3} \rho_{\rm f}=\frac{4\pi M_{\rm Pl}^2}{H_{\rm f}},
	\label{eq05}	
\end{equation}
we can relate $R$ to $M$. 
From now on, we identify $M\equiv M_{H_{\rm f}}$.

As a first step,  we derive the PBH mass $M$ as a function of the comoving wave
number $q_{\rm f}$ at the PBH formation time.
Combining the entropy conservation $S=C_1 g_* T^3 a^3$ = const
with radiation energy density 
$	\rho_{\rm rad} =C_2 g_* T^4 $,
we find
\begin{equation}
	\frac{\rho_{\rm rad,f}}{\rho_{\rm rad,0}}=\left(\frac{g_{*,0}}{g_{*,{\rm f}}}\right)^{1/3} \left(\frac{a_0}{a_{\rm f}}\right)^4,
	\label{eq04}
\end{equation}
where the quantity $g_*$ is the time-dependent number of effective massless degrees of freedom.
Then, from the first Friedmann equation
it follows  that the Hubble rate $H_{\rm f}$ at the formation time 
is related to the Hubble rate  $H_0$ at present as
\begin{equation}
	H_{\rm f}=H_0 \Omega_{\rm rad,0}^{1/2} (a_0/ a_{\rm f})^2 (g_{*,0}/g_{*,{\rm f}})^{1/6}, 
	\label{eq01}
\end{equation}
where $\Omega_{\rm rad,0}=\rho_{\rm rad,0}/\rho_{\rm crit}$,
with $\rho_{\rm crit}=3M_{\rm Pl}^2H_0^2$.
Combining this with
the horizon-crossing relations
$q_{\rm f}=a _{\rm f} H_{\rm f}$ and $q_0=a _0 H_0$,
we find 
\begin{equation}
	\frac{a_{\rm f}}{a_0} =  \Omega_{\rm rad,0}^{1/2}\left(\frac{g_{*,0}}{g_{*,{\rm f}}}\right)^{1/6}
	\frac{q _0}{q _{\rm f}} .
	\label{eq02}
\end{equation}
Next, substituting $H_{\rm f}$  
from (\ref{eq01}) and $a_{\rm f}$ from (\ref{eq02}) in $M = 4\pi M_{\rm Pl}^2/H_{\rm f}$, we find 
\begin{equation}
	M=\frac{4\pi M_{\rm Pl}^2}{ H_0}\Omega_{\rm rad,0}^{1/2}
	\left( \frac{g_{*,0}}{g_{*,{\rm f}}}\right)^{1/6} 
	\left(\frac{q _0}{q _{\rm f}}\right)^2 =1.90\cdot 10^{15}  
	\left(\frac{q _{\rm CMB}}{q _{\rm f}}\right)^2 M_\odot.
	\label{eq15}
\end{equation}
For the numerical estimate in (\ref{eq15}), we have used $M_{\rm Pl}=2.17645/\sqrt{8\pi}\times 10^{-8}$ kg, 
$\Omega_{\rm rad,0}=10^{-5}$,
the  Hubble radius
today $1/H_0=1.2\cdot 10^{26}$ m,  
$g_{*,{\rm f}}=106.75$, $g_{*,0}=3.36$ (see, e.g., Kolb and Turner \cite{kolb}), 
%the Planck length $l_{\rm Pl}=1.61624\sqrt{8\pi}\times 10^{-35}$ m, 
and the solar mass $M_{\odot}=2\cdot 10^{30}$ kg.

Finally, using  (\ref{eq15}) with $a_0=1$ we find 
\begin{equation}
	R\equiv \frac{1}{q_{\rm f}}=\frac{M^{1/2}}{(4\pi H_0)^{1/2}M_{\rm Pl}}
	\left( \frac{g_{*,{\rm f}}}{g_{*,0}}\right)^{1/12}\Omega_{\rm rad,0}^{-1/4}.
\end{equation}
It is convenient to express $R$ as a function of $M/M_{\odot}$,
\begin{equation}
	R= 1.4163\cdot 10^{16}\left(\frac{M}{M_{\odot}}\right)^{1/2} {\rm m}
	=4.590\cdot 10^{-7}\left(\frac{M}{M_{\odot}}\right)^{1/2} {\rm Mpc} .
	\label{eq0011}
\end{equation}

.

\subsubsection*{Estimate of the mass range}

Suppose the spectrum is significant between $q_{\rm min}$ and $q_{\rm max}$
as in the toy model (\ref{eq0014}). Inspired by the Tachyon model
inflation spectrum depicted in Fig.\ \ref{fig9} we use
$q_{\rm min}=10^4 q_{\rm CMB}$, $q_{\rm max}=10^{16} q_{\rm CMB}$.
Then, the BH production will be suppressed for
$R>R_{\rm max}=1/q_{\rm min}$ and  $R<R_{\rm min}=1/q_{\rm max}$.
The masses corresponding to  $R_{\rm min}$ and $R_{\rm max}$ are
\begin{equation}
	M_{\rm min}=\left(\frac{R_{\rm min}}{4.59\cdot 10^{-7}}\right)^2
	\simeq 1.91\cdot 10^{-17} M_{\odot},
	\label{eq0016}
\end{equation}
\begin{equation}
	M_{\rm max}=\left(\frac{R_{\rm max}}{4.59\cdot 10^{-7}}\right)^2
	\simeq 1.91\cdot 10^7 M_{\odot}.
	\label{eq0017}
\end{equation}
Then, according to (\ref{eq0015}), the function $\sigma(M)$ will  have an
approximate top hat profile
with $M$ ranging from $M_{\rm min}$ to $M_{\rm max}$ and the top value
$\sigma_{\rm max}$ defined by (\ref{eq0018}).

\subsection{PBH mass fraction}
\label{fraction}
The density of PBHs of mass $M$ at the formation time is a fraction $\beta(M)$ of the total background density $\rho_{\rm rad,f}$, 
\begin{equation}
	\rho_{\rm PBH,f}=\beta(M)\rho_{\rm rad,f} .
\end{equation}
We now describe two ways of calculating $\beta(M)$ that 
appear in the literature: The Press-Schechter and the Critical collapse and peaks (CCP) formalism, 
and compare the two approaches using the toy-model spectrum.

\subsubsection{Press-Schechter (PS) formalism}

In the Press-Schechter formalism of gravitational collapse, the mass fraction
%$\beta(M)$
of PBHs of mass $M$ is described by the probability that the overdensity $\delta$ is above a certain
threshold value $\delta_c$ for collapse. Assuming $\delta$ is a Gaussian random variable with mass (or
scale) dependent variance, the mass fraction $\beta(M)$ at the time of formation is then given by \cite{sasaki,bhaumik,yogesh}
\begin{equation}
	\beta(M)=\frac12 {\rm erfc}\left(
	\frac{\delta_c}{\sqrt2 \sigma}\right)
	\equiv\frac{1}{\sqrt{2\pi}\sigma}\int_{\delta_c}^{\infty}
	d\delta \exp \left(-\frac{\delta^2}{2\sigma^2}\right) .
\end{equation}
This function can be approximately expressed in terms of elementary functions by
\cite{wolfram}
\begin{equation}
	\beta(M)\simeq\frac{\exp \left(-\delta_c^2/(2\sigma^2)\right)}{\sqrt{\pi}
		\delta_c/(\sqrt2 \sigma)+\sqrt{\pi\delta_c^2/(2 \sigma^2)+4}} .
	\label{eq0006}
\end{equation}

\subsubsection{Critical collapse and peaks (CCP) formalism}

Given linear density contrast  $\delta$, the PBH mass can be well approximated 
by the scaling law for critical collapse \cite{niemeyer,young}
\begin{equation}
	M_{\rm PBH}(\delta)=K M (\delta_m(\delta)- \delta_c)^{\gamma} ,
	\label{eq0003}
\end{equation}
where $\delta_m$ is the smoothed density contrast, $M\equiv M_{H_{\rm f}}$ is the mass within the cosmological horizon at the PBH formation time $t_{\rm f}$ in radiation era, and
$\gamma\simeq 0.36$ is the critical exponent at $t_{\rm f}$
\cite{koike}.
The parameter $ K\simeq 4$ depends on the specific profile   of the
collapsing overdensity. 
The smoothed density contrast $\delta_m$  may be expressed in terms of $\delta$ as \cite{young}
\begin{equation}
	\delta_m=\delta -\frac38 \delta^2.
	\label{eq0004}
\end{equation}
Then, the PBH mass fraction is \cite{young} 
\begin{equation}
	\beta(M)=\int_{\delta_-}^{4/3}d\delta \frac{M_{\rm PBH}(\delta)}{M}N(\delta) ,
	\label{eq0105}
\end{equation}
where 
\begin{equation}
	N(\delta)=\frac{1}{4\pi^2}\left(\frac{\mu\delta}{\sigma^2}\right)^3
	\exp \left(-\frac{\delta^2}{2\sigma^2}\right)
\end{equation}
and the lower integral bound
\begin{equation}
	\delta_-=\frac{4}{3}\left(1-\sqrt{1-\frac32 \delta_c}\right)
\end{equation}
is the smaller root of the quadratic equation 
\begin{equation}
	\delta -\frac38 \delta^2-\delta_c=0.
\end{equation}
Then, using (\ref{eq0003}) with $K=4$ and (\ref{eq0004}), we 
obtain 
\begin{equation}
	\beta(M)=\frac{1}{\pi^2}\frac{\mu^3}{\sigma^7}\int_{\delta_-}^{4/3}d\delta
	\left(\delta -\frac38 \delta^2-\delta_c\right)^\gamma \delta^3 
	\exp \left(-\frac{\delta^2}{2\sigma^2}\right) .
	\label{eq0005}
\end{equation}

\subsubsection{Comparison}
\label{comparison}

The CCP formalism is generally considered to be more refined 
\cite{young5,cole} compared with the PS.
Whereas the PS formalism has one universal condition that the density must be above the threshold value, the CCP formalism introduces an additional constraint: PBHs form at peaks of the density \cite{young3,young}. 
In particular, as pointed out in Ref. \cite{kitajima}, the wide variety of the threshold value
implies the difficulty in the PS approach: the PBH formation criterion is not
a simple matter of the universal threshold for a one-point value of some averaged random
field, but we need to deal with the spatial dependence of the local overdensity profile. The two methods have been compared in a comprehensive recent analysis by Yoo et al.\ \cite{yoo} (see also Pi et al.\ \cite{pi}) that takes into account the mutually related issues of PBH formation criterion, statistical treatment of non-linear variables, and use of a window function.

To compare these two methods quantitatively, we can use the toy-model spectrum (\ref{eq0014}). In this model, the function $\beta(M)$ may be approximated by a similar top-hat profile
\begin{equation}
	\beta(M)=\beta_0 \Theta(M-M_{\rm min})\Theta(M_{\rm max}-M),
	\label{eq0019}
\end{equation}
where the  height $\beta_0$ can be estimated using either (\ref{eq0006}) or (\ref{eq0005}).
By way of example, we will use $\gamma=0.36$, $\delta_c=0.56$, and
$\sigma=(4/9)\sqrt{\mathcal{P}_0/2}$ with $\mathcal{P}_0=0.2$ or $0.4$. Then, using (\ref{eq0005}) we obtain  
$\beta_0=1.536\cdot 10^{-7}$  
%	$\beta_0=1.53592674048\cdot 10^{-7}$  	
for $\mathcal{P}_0=0.2$ and 
$3.429\cdot 10^{-5}$ 
%	$3.42927182895\cdot 10^{-5}$ 	
for $\mathcal{P}_0=0.4$. In contrast, using 
(\ref{eq0006})
we obtain $\beta_0=3.441\cdot 10^{-5}$ and  $2.489\cdot 10^{-3}$
%	we obtain $\beta_0=3.4408\cdot 10^{-5}$ and  $2.4893\cdot 10^{-3}$
for $\mathcal{P}_0=0.2$ and $\mathcal{P}_0=0.4$, respectively.
	Hence, for a broad power spectrum similar to the toy-model spectrum (\ref{eq0014}), one may expect that the PS prediction based on Eq.\ (\ref{eq0006}) will be by about a factor of 100 larger than the prediction of the CCP formalism based on Eq.\ (\ref{eq0005}).
	
	For the reasons stated above, in the following, we will use the
	more conservative CCP formalism. However, it is fair to mention that the quantitative difference
	between the two methods could  be reduced by an effective
	calibration — for example, by adjusting the collapse threshold $\delta_{\rm c}$ or redefining 
	$\sigma$ so that PS and CCP results become more consistent. 
	In our example above, the height of the mass fraction $\beta_0$ decreases very quickly with decreasing $\mathcal{P}_0$, to wit, with decreasing $\sigma$. Hence, using different window functions in the two approaches and thereby redefining $\sigma$ could reduce disagreement between PS and CCP methods. 
	In a recent study \cite{pi}, it has been shown that incorporating a smoothing procedure into the peak theory framework  alleviates 
	the tension between peaks theory and the
	Press–Schechter formalism.

\subsection{PBH abundance} 
The abundance of PBHs  with mass in the interval $(M,M + dM)$ related  to the total dark matter (DM) today is defined as 
\cite{sasaki,young,inomata,ballesteros,byrnes,zhang}
\begin{equation}
	\frac{d\Omega_{\rm PBH,0}}{\Omega_{\rm DM,0}} 
	=f_{\rm PBH} d\ln M .
	\label{eq8001}
\end{equation}
Here, $f_{\rm PBH}$  is the mass-dependent PBH fraction defined as
\begin{equation}
	f_{\rm PBH}\equiv 
	\frac{\rho_{\rm PBH,0}}{\Omega_{\rm DM,0}\rho_{\rm crit}}
	=\frac{\rho_{\rm PBH,f}}{\Omega_{\rm DM,0}\rho_{\rm crit}}\left(\frac{a_{\rm f}}{a_0}\right)^3=
	\frac{\beta(M)\rho_{\rm rad,f}}{\Omega_{\rm DM,0}\rho_{\rm crit}}\left(\frac{a_{\rm f}}{a_0}\right)^3 ,
\end{equation}
where $\Omega_{\rm DM,0}$ is the fraction of DM today.
We have taken into account that PBHs behave like matter, 
i.e., $\rho_{\rm PBH,0}a_0^3=\rho_{\rm PBH,f} a_{\rm f}^3$ and that the PBH density at the formation time is a fraction $\beta(M)$ of the total background density $\rho_{\rm rad,f}$.
Substituting $\rho_{\rm rad,f}$ from entropy conservation (\ref{eq04}),  $a_{\rm f}/a_0$ from (\ref{eq02}), and $q _{\rm f}/q _0$ from (\ref{eq15}),
we obtain
\begin{equation}
	f_{\rm PBH}=\frac{\beta(M)\Omega_{\rm rad,0}^{3/4}}{\Omega_{\rm DM,0}}
	\left(\frac{g_{*,0}}{g_{*,{\rm f}}}\right)^{1/4}\left(\frac{M_0}{M}\right)^{1/2} ,
\end{equation}
where 
\begin{equation}
	M_0=4\pi M_{\rm Pl}^2/H_0
\end{equation}
is the mass of the present cosmological horizon.
If we use $g_{*,0} = 3.36$, $\Omega_{\rm rad,0} = 10^{-5}$, $\Omega_{\rm DM,0}=0.265$,
$H_0=70\;\rm{km\, s^{-1}Mpc^{-1}}$, and $M_\odot = 2\cdot 10^{30}$ kg, we find 
\begin{equation}
	f_{\rm PBH}=\frac{\beta(M)}{1.67\cdot 10^{-8}}\left(\frac{M_\odot}{M}\right)^{1/2}
	\left(\frac{106.75}{g_{*,{\rm f}}}\right)^{1/4} .
	\label{eq5000}
\end{equation}
\begin{figure}[ht]
	\centering
	\includegraphics[width=0.9\textwidth ]{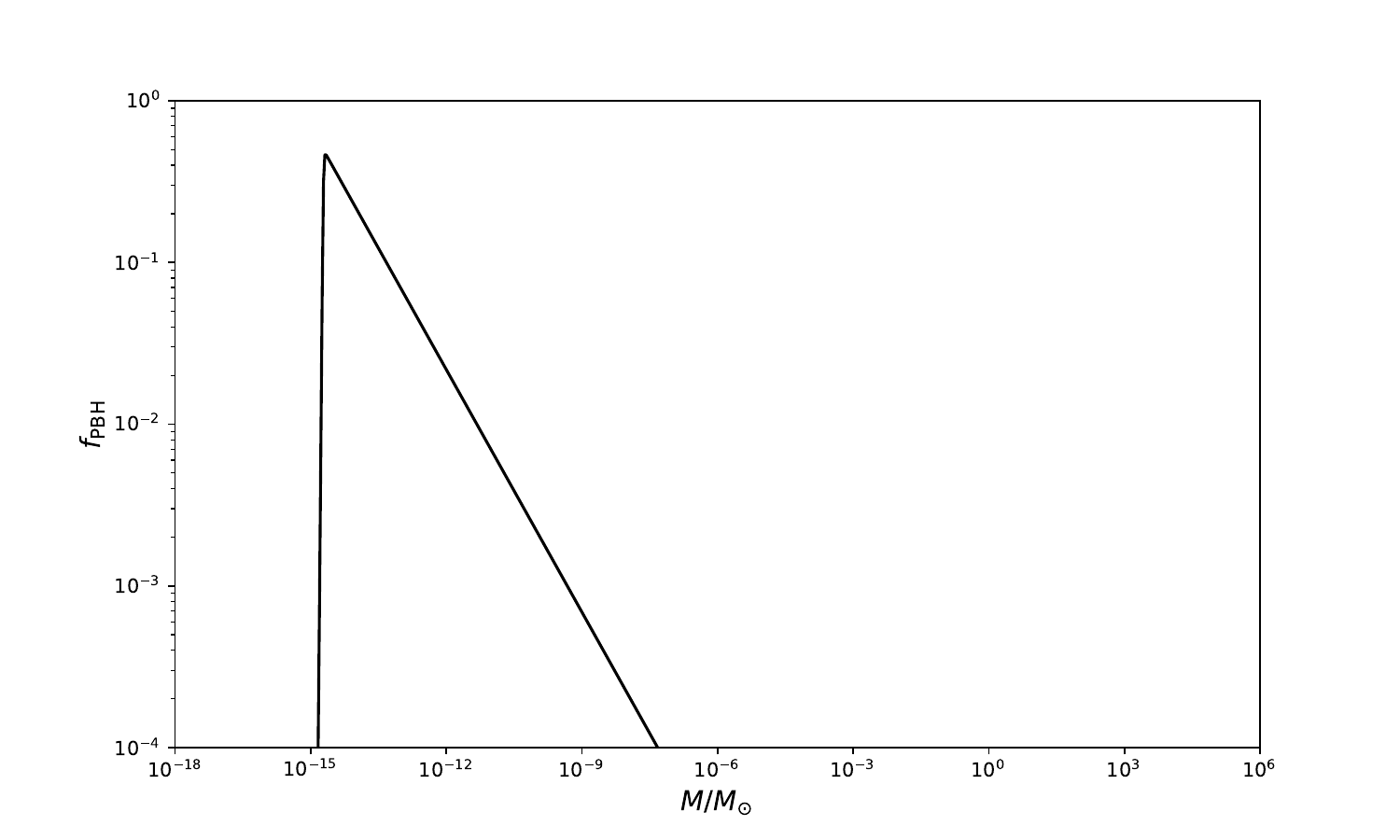}
	\caption{The fraction of PBH dark matter versus $M$ for the flat power spectrum (\ref{eq0014})
		with $q_{\rm min}=10^{4}q_{\rm CMB}$, $q_{\rm max}=10^{15}q_{\rm CMB}$, and $\mathcal{P}_0=0.08866$.
		%gnuplot
		%plot [10**(-20):10**10] [10**(-4):2.*10**(-1)] (0.5+0.5*tanh(1.*10**18*(x-1.9*10**(-17))))*(7.44924512719/1.67)*10**(-10)/sqrt(x)
		%gnuplot, new figure 16.5.2025
		%plot [10**(-20):10**10] [10**(-4):2.*10**(+0)] (0.5+0.5*tanh(1.*10**16*(x-1.9*10**(-15))))*(3.724622563595/1.67)*10**(-8)/sqrt(x)
		%gnuplot, new figure 18.5.2025
		%plot [10**(-20):10**10] [10**(-4):2.*10**(+0)] (0.5+0.5*tanh(1.*10**16*(x-1.91*10**(-15))))*(3.6537/1.67)*10**(-8)/sqrt(x)
	}	
	\label{fig14}
\end{figure}

\subsubsection*{PBH abundance in the top-hat model} 

For an estimate based on the top-hat toy model,  we  use 
$\beta(M)$ defined by (\ref{eq0019})
with mass range $(M_{\rm min},M_{\rm max})$ 
defined by (\ref{eq0016}) and (\ref{eq0017}) 
and $\beta_0$ calculated in Sec.~\ref{fraction} for a chosen $\mathcal{P}_0$.
Obviously, the values $\mathcal{P}_0$ and $q_{\rm max}$ (or the corresponding $M_{\rm min}$) are decisive since
$\beta(M)$ is extremely sensitive to $\mathcal{P}_0$ and since $f_{\rm PBH}$ 
peaks just above $M_{\rm min}$.

As an example, we take  
$q_{\rm max}=10^{15}q_{\rm CMB}$ corresponding to $M_{\rm min}=1.91\cdot 10^{-15}M_\odot$  
and we use the approximate 
expression (\ref{eq0019}) with $\beta_0$ estimated using (\ref{eq0005}). 
Then, for 
$\mathcal{P}_0=0.2$ 
we find $\beta_0=1.536\cdot 10^{-7}$  and
$f_{\rm PBH,peak}=2.105\cdot 10^8$,
whereas for $\mathcal{P}_0=0.08866$ we obtain $\beta_0=3.6537\cdot 10^{-16}$  
and
$f_{\rm PBH,peak}=0.50061$. 
In Fig.\ \ref{fig14}, we plot the fraction of PBH dark matter as a function of  $M$ 
at the formation time for the top-hat power spectrum (\ref{eq0014})
with height $\mathcal{P}_0=0.08866$. 
The choice of 
$q_{\max}$ in the top-hat profile is crucial since the position $M_{\rm min}$ of the
DM fraction peak is proportional to $1/q_{\rm max}^2$
(Eq.\ (\ref{eq0016})). 
There exists a natural cutoff at 
about $q_{\rm max}\sim 10^{17}q_{\rm CMB}$
which corresponds to $M_{\rm min}\simeq 10^{-18} M_\odot\simeq 10^{15}$ g. 
The PBHs of masses smaller than this would have evaporated completely until today.

\subsubsection*{PBH abundance in the PLLS and Tachyon models} 

\begin{figure}[ht]
	\centering
	\includegraphics[width=\textwidth ]{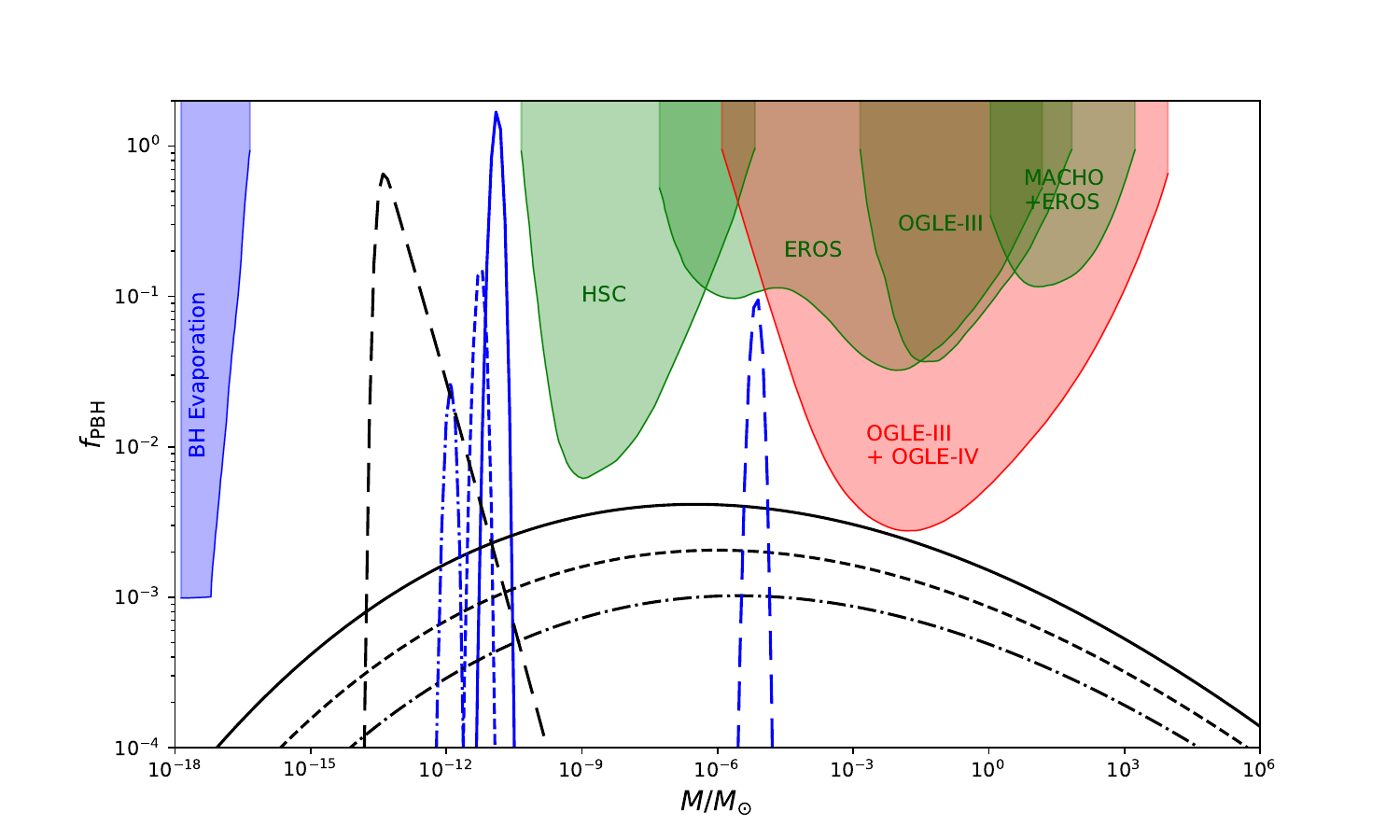}
	\caption{The fraction of PBH dark matter versus the PBH mass $M$  in the PLLS model (blue lines) and the Tachyon model (black lines) together with observational constraints. 
		The corresponding input parameters and initial values are listed in Table \ref{table1}.
		The green contours mark the limits
		determined by  EROS \cite{tisserand}, OGLE-III \cite{wyrzykowski}, Hyper Suprime-Cam (HSC)
		\cite{niikura}, and MACHO+EROS \cite{blaineau}.
		The orange contour marks the constraints by OGLE-III + OGLE-IV \cite{mroz}.
	}
	\label{fig15}
\end{figure}

In Fig.\ \ref{fig15},  we plot the fraction of PBH dark matter as a function of  $M$ 
for the power spectra  approximated by 
equation
(\ref{eq5021}) for both the PLLS and the
Tachyon models.
The theoretical curves correspond to the
power spectra calculated using specifically tuned input parameters listed in Table \ref{table1}.
The requirement that the PBH abundance is outside the regions constrained by observations dictates the choice of input parameters.
\begin{table}%[b]
	\centering
	\caption{Input parameters corresponding to $f_{\rm PBH}(M)$ plotted in Fig.\ \ref{fig15}}
	\begin{tabular}{||c|c|c|c|c|c|c|c||}
		\multicolumn{8}{c}{}  \\
		%	\hline
		%\begin{tabular}{|c|l|l|l|l|l|l|c|}
		\hline
		\textbf{Model} & $V_0 [M_{\rm Pl}^4]$ & $\lambda\times 10^{6}$ & $\phi_0 [M_{\rm Pl}]$ & $\eta_{\text{in}}$ & $\phi_{\text{in}} [M_{\rm Pl}]$ &
		$\delta_c$ &
		Line Style\\
		%\noalign{\hrule height 1.2pt}
		\hline
		
		& 			& 7.5490 & 0.835370 & $-1.4500 \times 10^{-8}$ & 5.2200 &	   & full        \\ \cline{3-6}	\cline{8-8}
		\textbf{PLLS}  & $10^{-16}$ & 7.5400 & 0.835370 & $-2.8174 \times 10^{-7}$ & 5.2500 & 0.51 & dashed  \\\cline{3-6}\cline{8-8}
		$\alpha = 1.5$ & 			& 7.5400 & 0.835370 & $-2.8174 \times 10^{-7}$ & 5.3000 & 	   & dash-dotted  \\\cline{3-6}\cline{8-8}
		& 			& 3.9523 & 0.895870 & $-1.0089 \times 10^{-3}$ & 5.1124 & 	   & long dashed  \\
		%\noalign{\hrule height 1.2pt}
		\hline
		
		& &  		 & 		 &   				   & 1.1000    & 		& full         	   \\\cline{6-6}\cline{8-8}
		\textbf{Tachyon} & $10^{-16}$ & 7.5020 & 0.259500 & $-6.1200 \times 10^{4}$ & 1.1005 & 0.56 & dashed  \\\cline{6-6}\cline{8-8}
		& &       &        &                     & 1.1010    & 		& dash-dotted  \\\cline{8-8}\cline{3-6}
		& &	 7.6000 &  0.259831 & $ -5.3625 \times 10^{4}$ & 1.0520 & 		 	& long dashed  \\
		\hline
	\end{tabular}
	\label{table1}
\end{table}

The critical collapse thresholds $\delta_{\rm c}$ are calculated for each model using the procedure described in appendix \ref{threshold}.
The variations of input parameters and initial values in the ranges given in Table \ref{table1}  do not substantially affect the shape of the spectra, so the variations of
the corresponding critical collapse thresholds are insignificant.
The obtained values  are
$\delta_{\rm c}=0.51$ for the PLLS model 
and $\delta_{\rm c}=0.56$ for the Tachyon model spectra.
The fraction of PBH is calculated using Eqs.\  (\ref{eq0005}) and  (\ref{eq5000}), with $\sigma$ and $\mu$ obtained from
(\ref{eq0012}) and (\ref{eq0013}), respectively.
The data for the constraints are from Ref.\ \cite{mroz}. 
\begin{figure}[t!]
	\begin{center}
%		\includegraphics[width=0.48\textwidth,trim= 0 0cm 0 0cm]{fig5-v1}
%		\hspace{0.02\textwidth}
%		\includegraphics[width=0.48\textwidth]{fig8-v1}
		\includegraphics[width=0.49\textwidth]{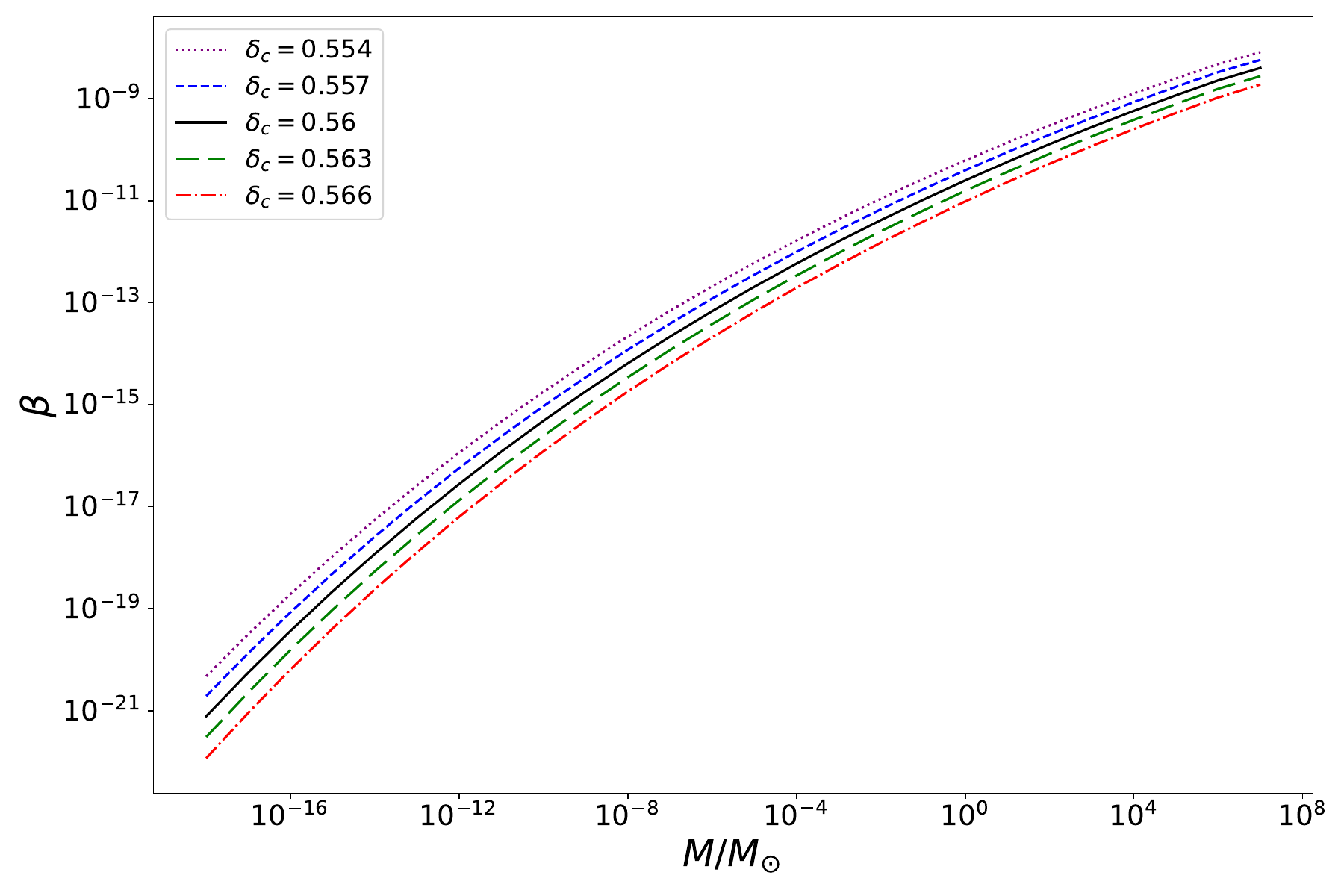}
        \includegraphics[width=0.49\textwidth]{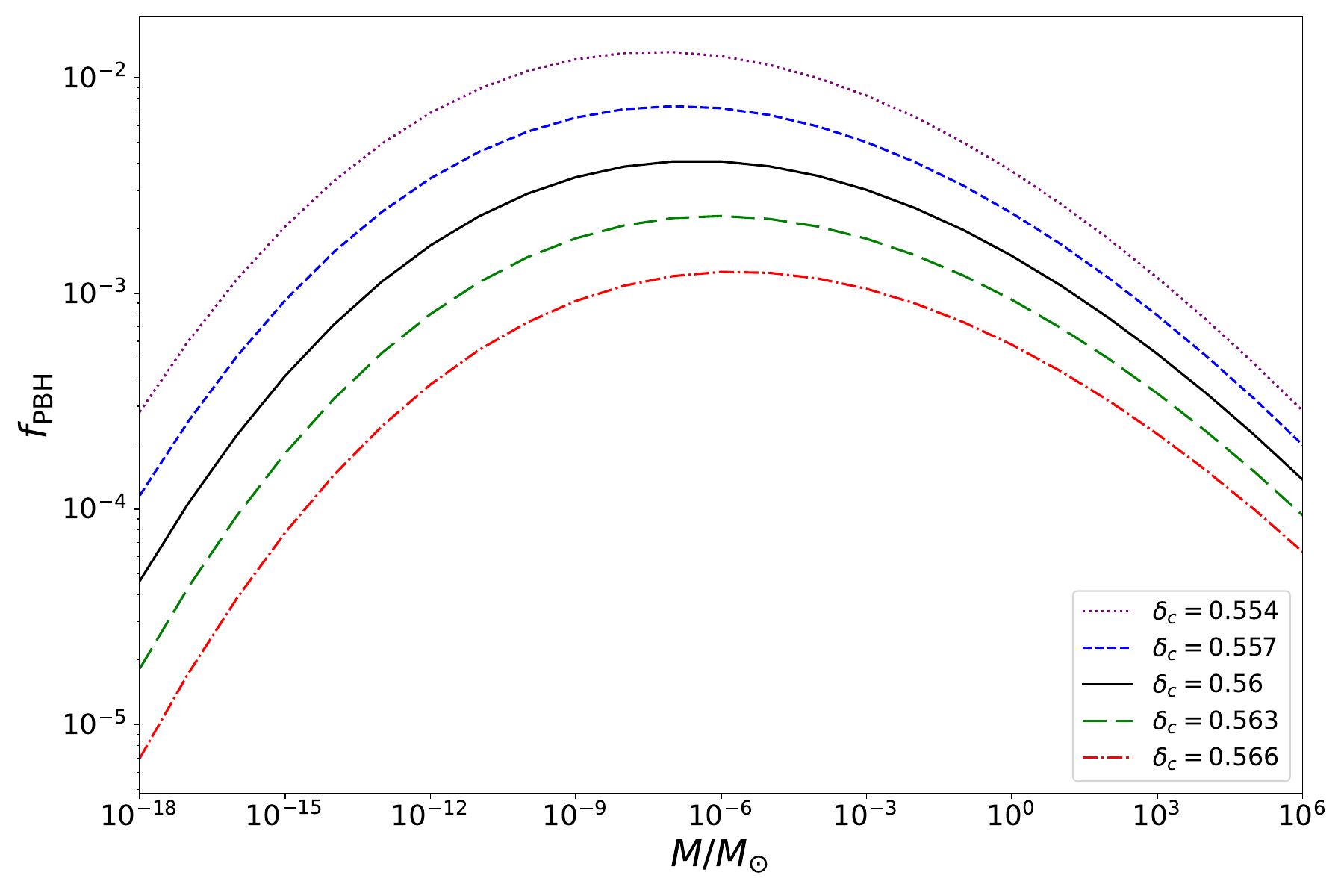}
		\caption{The mass fraction $\beta$ (left panel) and fraction of PBH dark matter (right panel) versus $M$
			in the Tachyon model for various critical thresholds $\delta_{\rm c}$ and fixed set of parameters as in Table \ref{table2} with $\phi_{\rm in}=1.1 M_{\rm Pl}$.
		}
	
		\label{fig1}
	\end{center}
\end{figure}

The PLLS model with the set of parameters in Table \ref{table1} predicts the PBH fraction peaks in the observationally allowed windows of masses between $10^{-13}$ and $10^{-10}$ and between $10^{-6}$ and $10^{-4}$ solar masses. The Tachyon model, with the set of parameters in Table \ref{table1}, predicts experimentally allowed PBH abundance today in a broad mass range from $10^{-16}$ to $10^{6}$ solar masses, except for the special case (long dashed black line on Fig.\ \ref{fig15}) 
with the allowed masses between $10^{-14}$ and $10^{-10}$ solar masses.  

For a DM prediction, we need the integrated PBH dark matter fraction. To this end, we 
use the definition (\ref{eq8001}) so the integrated PBH dark matter abundance in the  mass interval $(M_{\rm min},M_{\rm max})$ related to the total DM fraction is given by
\begin{equation}
	F_{\rm PBH}=
	\int_{M_{\rm min}}^{M_{\rm max}}f_{\rm PBH} \frac{d M}{M} .
\end{equation}
The results for both models are given in Table \ref{table3} for each set of parameters
listed in Table \ref{table1}. 
The PBH contribution to DM today could be significant 
and,  in particular cases, even sufficient to account for the total DM. However, it is important to stress that the parameterization of the
Tachyon model that leads to $F_{\rm PBH}=0.978$ corresponds to the spectrum
not supported by the CMB observational data in the vicinity of the CMB pivot scale $q_{\rm CMB}$.
\begin{table}[t]
	\centering
	\caption{Integrated PBH DM fractions corresponding to $f_{\rm PBH}(M)$ plotted in Fig.\ \ref{fig15}}
	{\small
		\begin{tabular}{||c|c|c||}
			\multicolumn{3}{c}{}  \\
			%	\hline
			%\begin{tabular}{|c|l|l|l|l|l|l|c|}
			\hline
			\textbf{Model} &
			$F_{\rm PBH}$ [\%] &
			Line Style
			\\
			%\noalign{\hrule height 1.2pt}
			\hline
			
			&  100 	& full blue     \\ \cline{2-3}
			\textbf{PLLS} & 9.0 & dashed blue \\\cline{2-3}
			$\alpha = 1.5$& 1.4 & dash-dotted blue  \\\cline{2-3}
			& 6.2	& long dashed blue  \\
			%\noalign{\hrule height 1.2pt}
			\hline
			
			& 10.6     	& full black  	   \\ \cline{2-3}
			\textbf{Tachyon}& 5.3 & dashed black  \\ \cline{2-3}
			& 2.6	& dash-dotted black  \\ \cline{2-3}
			& 97.8	& long dashed black \\
			\hline
		\end{tabular}
	}
	\label{table3}
\end{table}
\subsubsection{Sensitivity to $\delta_{\rm c}$}
\label{sensitivity}
	The mass fraction $\beta$ and consequently the PBH abundance calculated in both PS and CCP formalisms is exponentially sensitive
	to the critical collapse threshold $\delta_{\rm c}$. 
	The
	value of $\delta_c$ depends on the shape of the collapsing curvature
	power spectrum \cite{musco2} (see also \cite{escriva,bhaumik,dalianis,cole,harada,musco3,kalaja,escriva2,stamou}). 
	In appendix 
	\ref{threshold}, we outline the calculation procedure. For a broad power
	spectrum that could be approximated by (\ref{eq0014}), we recover  $\delta_c\sim 0.56$ of Ref.\ \cite{musco2}. 
	For the PLLS and Tachyon models, 
	the variations of input parameters and initial values in the interesting ranges do not substantially affect the shape of the spectra. The variations of
	the corresponding critical collapse thresholds are within  
	$\delta_{\rm c}=0.51$ for the PLLS model 
	and $\delta_{\rm c}=0.56$ for the Tachyon model spectra.

	In Fig.\ \ref{fig1} we plot the mass fraction $\beta$ and the PBH abundance $f_{\rm PBH}$ in the Tachyon model for various $\delta_c$ 
	within the plausible uncertainty departure of about 1\% from the numerical value calculated in appendix \ref{threshold}.

\subsection{Comment on primordial non-Gaussianity}
\label{nongaussianity}
The three-point
correlation function 
\begin{equation}
	\langle\hat{\zeta}_{q_1}\hat{\zeta}_{q_2}\hat{\zeta}_{q_3}\rangle=
	(2\pi)^3 \delta(\mbox{\boldmath $q$}_1+\mbox{\boldmath $q$}_2+\mbox{\boldmath $q$}_3)
	f_{\rm NL} F(q_1,q_2,q_3)
	\label{eq0066}
\end{equation}
provides the prime diagnostic of non-Gaussianity of inflationary fluctuations \cite{baumann}.
The quantity $f_{\rm NL}$ is a dimensionless parameter defining the amplitude of non-Gaussianity and
the function $F$ captures the momentum dependence.

The lowest-order non-Gaussian corrections to the primordial power spectrum arise from a single
one-loop diagram, which leads to
\cite{byrnes2,unal,wang2}
\begin{equation}
	\mathcal{P}_{\rm S}(q)=\mathcal{P}_{\rm S,G}(q)+f_{\rm NL}^2 q^3
	\int \frac{d^3k}{4\pi}\frac{\mathcal{P}_{\rm S,G}(k)}{k^3}
\frac{\mathcal{P}_{\rm S,G}
(|\mbox{\boldmath $q$}-\mbox{\boldmath $k$}|)}{|\mbox{\boldmath $q$}-\mbox{\boldmath $k$}|^3},
\end{equation}
where $\mathcal{P}_{\rm S,G}$ is the dimensionless
power spectrum defined in (\ref{eq4024}), and the second term on the right-hand side is a non-Gaussian contribution which can be  rewritten as 
\begin{equation}
	\mathcal{P}_{\rm S,NG}(q)=f_{\rm NL}^2 q^2
	\int_0^\infty \frac{dk}{k^2}\mathcal{P}_{\rm S,G}(k)
	\int_{|q-k|}^{q+k}\frac{dk'}{{k'}^2}	\mathcal{P}_{\rm S,G}(k') .
	\label{eq3001}
\end{equation}

\"Unal \cite{unal} estimated (\ref{eq3001}) for the  
log-normal gaussian spectrum 
\begin{equation}
\mathcal{P}_{\rm S,G}=A\exp[-\ln^2(k/k_{\rm peak})/(2\sigma^2)].
\label{eq1062}
\end{equation}
 For an amplitude $A=0.01$, and $f_{\rm NL}=3$, he found that the non-Gaussian term is subdominant by about a factor of 10. Clearly, if the amplitude were 10 times larger, the non-Gaussian term would be of the same order as the Gaussian one.
 Similarly, Riccardi, Taoso, and Urbano \cite{taoso} found that in realistic single-field inflationary models with ultra slow-roll,  by changing the peak amplitude of the curvature power spectrum by about a factor of two, one could
 obtain the same abundance calculated with the Gaussian approximation as the one 
 obtained in the presence of local non-Gaussianities.

In the $k$-essence models, the largest non-Gaussianity  peakes at the so-called {\em equilateral} configuration
with $q_1\sim q_2\sim q_3$.
The non-Gaussianity amplitude of an equilateral triangle  in a general
$k$-essence is given by \cite{baumann,chen}
\begin{equation}
	f_{\rm NL}^{\rm equil} = - \frac{35}{108} \left( \frac{1-c_{\rm s}^2}{c_{\rm s}^2} \right) 
	+ \frac{20}{81} \Lambda,
	\label{eq0067}
\end{equation}
where
\begin{equation}
	\Lambda \equiv  X^2\frac{ \mathcal{L}_{XX}^2-
		(1/3)\mathcal{L}_{X} \mathcal{L}_{XXX}}{ \mathcal{L}_{X}^2 + 2 X\mathcal{L}_{X} \mathcal{L}_{XX}} .
\end{equation}
The quantity $\Lambda$ in the Tachyon model is identically zero. 
%Now, we estimate the non-Gaussianity amplitude in our Tachyon model. 
As the sound speed deviates from unity  most at the end of the slow roll regime, we estimate the equilateral amplitude at the end of 
inflation, neglecting possible post-inflationary effects.
For an estimate 
we can use  $c_{\rm s}^2=1/3$,  given by (\ref{eq3039}) for $\varepsilon_1=1$, yielding
\begin{equation}
	f_{\rm NL}^{\rm Tachyon} \approx   \left. f_{\rm NL}^{\rm equil} \right|_{\varepsilon_1=1}= -\frac{70}{108} . 
	\label{eq0060}
\end{equation}
In the PLLS model, we find 
\begin{equation}
	\Lambda=\frac{1-3\alpha+2\alpha^2}{3+6\alpha} ,
\end{equation}
so for $\alpha=1.5$ we have
$\Lambda=1/12$. Thus, using (\ref{eq0067}) with $c_{\rm s}^2=1/2$, for the PLLS model
we find $f_{\rm NL}\approx f_{\rm NL}^{\rm equil}=-295/972$
\begin{equation}
	f_{\rm NL}^{\rm PLLS} \approx   \left. f_{\rm NL}^{\rm equil} \right|_{\varepsilon_1=1}= -\frac{295}{972} . 
	\label{eq1060}
\end{equation}
Both values in (\ref{eq0060}) and (\ref{eq1060})
are well within the observational constraints provided by the Planck 2018 collaboration \cite{planck2018}:
$ f_{\rm NL}^{\rm equil} =-26\pm 47$.

To estimate the NG contribution to the power spectrum in the PLLS model, we could approximate the spectrum of Fig.\ \ref{fig7} by the log-normal function (\ref{eq1062}) with an amplitude of about $A=0.1$. In our estimate,  $f_{\rm NL}\simeq 0.3$, so the NG contribution to the power spectrum would amount to a factor of  $(0.3/3)^2=0.01$ suppression compared to the
NG term in Ref.\ \cite{unal}. 
Since our amplitude is a factor of 10 larger than that of Ref.\ \cite{unal}, the NG contribution
enhances the PLLS power spectrum of Fig.\ \ref{fig7} by about 10\%.

The results obtained for the log-normal spectrum of Ref.\ \cite{unal} cannot be used for the broad spectrum of our Tachyon model. However, we can obtain a rough estimate by making use of the top-hat power spectrum (\ref{eq0014})).
In this model, we find
\begin{eqnarray}
	\mathcal{P}_{\rm S,NG}=f_{\rm NL}^2\mathcal{P}_0^2 q^2 
	\int_{q_{\rm min}}^{q_{\rm max}} \frac{dk}{k^2}
	\left[\frac{ \Theta(|q-k|-q_{\rm min})\Theta(q_{\rm max}-|q-k|)}{|q-k|}\right.
	\nonumber\\
	\left. -\frac{ \Theta(q+k-q_{\rm min})\Theta(q_{\rm max}-q-k)}{q+k}
	\right].
\end{eqnarray}
\\
The integration yields
\begin{equation}
	\mathcal{P}_{\rm S,NG}=f_{\rm NL}^2\mathcal{P}_0^2
	\left[
	2\ln\frac{q^2-q_{\rm min}^2}{q_{\rm min}^2}+
	2\ln\frac{q_{\rm max}-q}{q_{\rm max}}
	-\frac{2q^2}{q^2-q_{\rm min}^2}
%	-\frac{q}{q+q_{\rm min}}
	+\frac{q}{q_{\rm max}}+\frac{q}{q_{\rm max}-q} 	
	\right] .
\end{equation}
In the middle of the distribution, where $q_{\rm min}\ll q\ll q_{\rm max}$
we find an estimate
\begin{equation}
	\mathcal{P}_{\rm S,NG}\simeq 2f_{\rm NL}^2\mathcal{P}_0^2
\left(\ln \frac{q^2}{q_{\rm min}^2}-1 +\mathcal{O}(q_{\rm min}^2/q^2)
+\mathcal{O}(q^2/q_{\rm max}^2)\right) .
\end{equation}
For our estimated $f_{\rm NL}\simeq-\frac{70}{108}$ and the parameters of our top-hat model of Fig.~\ref{fig14} 
$\mathcal{P}_0=0.09$, $q/q_{\rm min}=10^4\div 10^8$, we find 
$\mathcal{P}_{\rm S,NG}=0.12\div 0.24$, to be compared with $\mathcal{P}_{\rm S,G}=0.09$.
Hence, the NG contribution in the top-hat model is positive and of the order $\mathcal{O}(1\div 2)\mathcal{P}_{\rm S,G}$.  Thus,  we expect the peak of the power spectrum that includes a non-Gaussian contribution in the Tachyon model to be enhanced by about a factor of $\mathcal{O}(2\div 3)$
compared with the Gaussian power spectrum. 

However, there is in addition an intrinsic non-Gaussianity owing to the nonlinear relation (\ref{eq3002}) between $\zeta$ and $\delta_r$ \cite{young,kawasaki,deluca2}. In the presence of such non-Gaussianity, the amplitude of the power spectrum can be
significantly reduced by a factor $\mathcal{O}(2\div 3)$ \cite{deluca2} compared to the Gaussian (linear) case, where one approximates (\ref{eq3002}) by the linear relation (\ref{eq3003}). 
In this way, the effect of this intrinsic non-Gaussianity works in the opposite direction compared to the primordial non-Gaussianity discussed above, so these two effects might partly annihilate. However, a precise estimate of the total outcome requires more accurate calculations, which go beyond the scope of the present paper.

Approximating the nonlinear relation (\ref{eq3002}) by the linear one can also affect the critical threshold $\delta_{\rm c}$.
According to Kehagias, Musco, and Riotto  \cite{kehagias}, the impact of the intrinsic non-Gaussianity on 
a relative change of $\delta_{\rm c}$  with respect
to the value obtained using a linear relation between $\zeta$ and $\delta_r$ is of the order of a few percent. 

\section{Summary and conclusions}
\label{conclude}
We have analyzed the PBH production in the early universe in
the top-hat toy model and two $k$-essence models of inflation. The analysis of the top-hat toy model reveals that
the Press-Schechter prediction for the PBH mass fraction $f_{\rm PBH}$ based on Eq.\ (\ref{eq0006}) is by a factor of 100 larger than the prediction of the CCP formalism based on Eq.\ (\ref{eq0005}). For this reason, we have employed the more conservative CCP formalism in calculating $f_{\rm PBH}$ for realistic power spectra.

Besides, the analysis of the top-hat model shows that	
the	PBH abundance is very sensitive to the height of the power spectrum. The top-hat  power spectrum with height equal to or above 0.2 
yields too large PBH abundance with $f_{\rm PBH} \gg 1$ whereas 
a height  below 0.08 yields a small peak value of $f_{\rm PBH}$
less than 0.02.

Regarding the PLLS and Tachyon models, we have demonstrated that it is possible 
to obtain a significant PBH DM fraction by fine-tuning the input parameters, $V_0$, $\phi_0$, $\lambda$, and initial conditions for the background equations. 
However,  the normalization requirement for the spectrum to fit the observational value at the pivot CMB scale constrains the tuning of these parameters. This normalization requirement can be easily implemented by utilizing the rescaling invariance \cite{bilic}.

The PBH abundance is fairly sensitive to the value of the critical collapse threshold 
$\delta_{\rm c}$.  We have found that the shape of the spectra in both models does not significantly change  
with slight variations in input parameters. As a consequence, the critical collapse threshold $\delta_{\rm c}$ is also not very sensitive to these variations.
Hence, the values of $\delta_{\rm c}$ calculated for the spectra presented in Figs.\ \ref{fig7} and \ref{fig9} for the PLLS and the Tachyon model, respectively, could have been applied for each set of parameters presented in Table \ref{table1}.
	However, as Fig.\ \ref{fig1} shows, the  relative change of $\delta_{\rm c}$ 
	(e.g., caused by the intrinsic non-Gaussianity \cite{kehagias}) within about $\pm$1\%
	could affect the PBH production by a factor of up to $4 \div 6$.

Our results for the PLLS model show that a suitable choice of input parameters different from those in Ref.\ \cite{papanikolaou} could yield a significant PBH abundance.
For a chosen set of parameters yielding the power spectrum of Fig.~\ref{fig7}, the PBH fraction has a few sharp peaks around the PBH
mass $M=10^{-11}M_\odot$. In contrast, the Tachyon model power spectrum of Fig.\  \ref{fig9}  predicts experimentally allowed PBH abundance today in a broad mass range (Fig.~\ref{fig15}).
In the PLLS model, the integrated PBH abundance of the highest peak (full blue line) in Fig.~\ref{fig15}  could account for total DM.  In the Tachyon model, the highest integrated contribution of PBHs to DM  estimate is about  $11\%$ (full black line), 
except for a particular parameterization (long dashed black line) where the integrated PBH contribution may reach up to $98\%$. However, the latter result should not be taken seriously since the spectrum corresponding to this parameterization is not supported by the observational data near the CMB pivot scale.

\appendix

\section{The critical collapse threshold} 
\label{threshold}
The critical collapse threshold $\delta_{\rm c}$ for PBH formation depends on the shape of the curvature perturbation spectrum. The shape parameter $\alpha$ describes the main features of
the profile in the length-scale interval $0 < r <r_m$ where PBHs form.
Here, we outline the procedure
for calculating $\delta_{\rm c}$ following  Musco \textit{et al.}\
\cite{musco2}.

The first step is to compute the value of the 
comoving length scale $\hat{r}_m$ of the perturbation related
to $r_m$ via the coordinate transformation $r=\hat{r}\exp{\zeta(\hat{r})}$.
To this end, we use the smoothed
power spectrum $\mathcal{P}_{\rm S}(q,\tau)$ 
defined as
\begin{equation}
	\mathcal{P}_{\rm S}(q,\tau)=\frac{2\pi^2}{q^3}\mathcal{P}_{\rm S}(q)T^2(q\tau) ,	
\end{equation}
where the transfer function $T(x)$ is defined as
\begin{equation}
	T(x)=3\frac{\sin(x/\sqrt3)-(x/\sqrt3)\cos(x/\sqrt3)}{(x/\sqrt3)^3}.
	\label{eq30}
\end{equation}
The conformal time interval $\tau$ is, according to (\ref{eq2}), approximately $\tau=-1/aH$. Now we assume that the scale $\hat{r}_m$ is a factor of 10 to 100
larger than the comoving horizon radius $1/(aH)$. Hence,  in Eq. (\ref{eq30}) we  may substitute $ -\hat{r}_m/p$ for $\tau$, where $p$ is a constant of order $10 \leq p\leq 100$ and, according to \cite{musco2}, the scale $\hat{r}_m$ is a positive root of the function
\begin{equation}
	I(x)=	\int_0^\infty \frac{dy}{y}\left[\left(y^2-1\right)
	\frac{\sin y}{y}+\cos y\right]\mathcal{P}_{\rm S}(y/x)T^2(y/p).
\end{equation}

As the next step, we compute 
the shape parameter $\alpha$. First, we find the  
linear Gaussian shape parameter $\alpha_{\rm G}$ using
\begin{equation}
	\alpha_{\rm G}= -\frac14 \left[	1+\dfrac{\int y dy 
		\cos y\,\mathcal{P}_{\rm S}(y/\hat{r}_m)T^2(y/p)}{
		\int dy 
		\sin y\,\mathcal{P}_{\rm S}(y/\hat{r}_m)T^2(y/p)	
	}
	\right].
\end{equation}
Then, we compute the shape parameter $\alpha$ by solving the
algebraic equation
\begin{equation}
	F(\alpha)(1+F(\alpha))\alpha=2\alpha_{\rm G} ,
	\label{eq31}
\end{equation}
where
\begin{equation}
	F(\alpha)=\left(1-\frac25\, 
	\frac{e^{-1/\alpha}\alpha^{1-5/(2\alpha)}}{f(\alpha)}
	\right)^{1/2},
\end{equation}
with
\begin{equation}
	f(\alpha)= \Gamma(5/(2\alpha))-\Gamma(5/(2\alpha),1/\alpha)
	\equiv\int_0^{1/\alpha} dt e^{-t} t^{5/(2\alpha)-1} .
\end{equation}

Finally, we compute the threshold
$\delta_{\rm c}$ as a function of $\alpha$.
Up to a few percent precision, the threshold can be expressed as an analytic function  
\begin{equation}
	\delta_{\rm c}(\alpha)=\frac{4}{15}\, 
	\frac{e^{-1/\alpha}\alpha^{1-5/(2\alpha)}}{f(\alpha)} .
	\label{eq33}
\end{equation}

\subsection*{Acknowledgments}
This work is supported by the ICTP-SEENET-MTP project NT-03 Cosmology-Classical and Quantum Challenges and the COST action CA1810 ``Quantum gravity phenomenology in the multi-messenger approach". The work of N. Bili{\' c} is supported by the COST Action CA21136 - Addressing observational tensions in cosmology with systematics and fundamental physics (CosmoVerse), while G. S. Djordjevi{\' c} acknowledges support by the COST Action CA23130 - Bridging high and low energies in search of quantum gravity (BridgeQG). 
%and CA21136 ``Addressing observational tensions in cosmology with systematics 
%and fundamental physics". 
%This article/publication is based upon work from COST Action CA21136 Addressing observational %tensions in cosmology with systematics and fundamental physics (CosmoVerse) supported by COST %(European Cooperation in Science and Technology).” 
N.\ Bili\'c is indebted to E.\ Saridakis for valuable comments.
M.\ Stojanovi\'{c} acknowledges the support by the Ministry of Science, Technological Development and Innovation of the Republic of Serbia under contract 451-03-137/2025-03/200113.
D.D.\ Dimitrijevi\'{c}, G.S.\ Djordjevi\'{c}, and M.\ Milo\v{s}evi\'{c} acknowledge the support by the same Ministry under contract 451-03-137/2025-03/200124.  G.S.\ Djordjevi\'c and 
and D.D.\ Dimitrijevi\' c
acknowledge the support by the CEEPUS Program RS-1514-03-2223 ``Gravitation and Cosmology". 
G.S.\ Djordjevi\'c acknowledges the hospitality of the CERN-TH.
N.\ Bili\'c acknowledges the hospitality of the Department of Physics, University of Ni\v{s}, where a part of his work has been completed.

\end{document}